\documentclass[conference]{IEEEtran}
\IEEEoverridecommandlockouts
\usepackage{cite}
\usepackage{amsmath,amssymb,amsfonts}
\usepackage{algorithmic}
\usepackage{graphicx}
\usepackage{textcomp}
\usepackage{xcolor}
\usepackage[lined,vlined,ruled]{algorithm2e}
\usepackage{subfigure}
\usepackage{epstopdf}
\usepackage{colortbl}
\usepackage{multirow}
\usepackage{enumitem}
\def\BibTeX{{\rm B\kern-.05em{\sc i\kern-.025em b}\kern-.08em
    T\kern-.1667em\lower.7ex\hbox{E}\kern-.125emX}}
\begin{document}

\title{Exploring Data and Knowledge combined Anomaly Explanation of Multivariate Industrial Data
}
\author{\IEEEauthorblockN{Xiaoou Ding\IEEEauthorrefmark{1}, Hongzhi Wang\IEEEauthorrefmark{2}, Chen Wang \IEEEauthorrefmark{4}, Zijue Li\IEEEauthorrefmark{2}, Zheng Liang \IEEEauthorrefmark{2}}
\IEEEauthorblockA{\IEEEauthorrefmark{1}\IEEEauthorrefmark{2}School of Computer Science and Technology, Harbin Institute of Technology.\\ \IEEEauthorrefmark{4}National Engineering Laboratory for Big Data Software, EIRI, Tsinghua University.\\
Email: \IEEEauthorrefmark{1}dingxiaoou@stu.hit.edu.cn, \IEEEauthorrefmark{2}\{wangzh,lizijue,lz20\}@hit.edu.cn, \IEEEauthorrefmark{4}wang\_chen@tsinghua.edu.cn}
}

\maketitle

\begin{abstract}
The demand for high-performance anomaly detection techniques of IoT data becomes urgent, especially in industry field. The anomaly identification and explanation in time series data is one essential task in IoT data mining. Since that the existing anomaly detection techniques focus on the identification of anomalies, the explanation of anomalies is not well-solved. We address the anomaly explanation problem for multivariate IoT data and propose a 3-step self-contained method in this paper. We
formalize and utilize the domain knowledge in our method, and
identify the anomalies by the violation of constraints. We propose set-cover-based anomaly explanation algorithms to discover the anomaly events reflected by violation features, and further develop knowledge update algorithms to improve the original knowledge set. Experimental results on real datasets from large-scale IoT systems verify that our method computes high-quality explanation solutions of anomalies. Our work provides a guide to navigate the explicable anomaly detection in both IoT fault diagnosis and temporal data cleaning.
\end{abstract}

\section{Introduction}
\label{sec1}
Anomaly is summarized as any unusual change in a value or a pattern, which does not conform to specific expectations \cite{DBLP:conf/kdd/ToledanoCBT17,DBLP:conf/kdd/DasuLS14}. Identifying the anomalies (roughly regarded as outliers, errors, and glitches) is one of the most challenging and exciting topics in data mining and data cleaning community~\cite{DBLP:conf/kdd/ToledanoCBT17, DBLP:journals/tkde/TakeuchiY06}. Researchers have gone a long way in anomaly detection studies in various fields. \cite{DBLP:series/synthesis/2014Gupta} introduces anomaly detection techniques for temporal data, which covers several kinds of temporal data including time series.  Anomaly detection techniques with different methods such as Density-, Window- and Constraint-based approaches have been developed and applied in various real scenarios (see \cite{DBLP:series/synthesis/2014Gupta,DBLP:journals/access/WangW20} as survey).

\indent The rapid development of sensor technologies and the widespread use of sensor devices witness the flowering of data management and data mining technologies in sensor data. The demand for high-performance mining techniques of Internet of Things (IoT) data also becomes urgent, especially in industry field. Time series is one of the most important types in IoT data \cite{DBLP:journals/access/WangW20,DBLP:journals/debu/DasuDS16}.
Thus, the anomaly identification and explanation tasks for time series data are essential. Despite of the advanced anomaly detection techniques, existing studies pay much attention to the \emph{identification} of anomalies and errors, leaving the anomaly \emph{reasoning} and \emph{explanation} not well solved. The characteristics of multivariate time series data in IoT applications urgently expect the detection methods to go further to explain the occurrence of anomalies, and thus, to achieve a higher dependability and interpretability in anomaly studies.

Anomaly explanation, or called explainable anomaly identification, will promote the involvement of domain knowledge in the techniques, and assist users to well understand the anomalies, which in turn improves the identification performance. It also complements existing data cleaning techniques considering the explanation discovery for errors and violations.
Considering the limitation of the current state-of-art anomaly explanation approaches, the challenges of the problem are as follows.
\newline \indent (1). \emph{Less attention paid to the dependency of data}.
Since that the \emph{multivariate} time series are collected with each sequence (\emph{i.e.}, attribute) from to one sensor, the abnormal data in sequences may not be completely independent. Anomalies would result in the glitches existing across multiple attributes with complex interactions. Treating all sequences independently may fail to correctly identify the actual errors in data.
\newline \indent (2). \emph{Under-utilized domain knowledge}. Discovering the interactions and causes of anomalies requires the involvement of \emph{knowledge} which comes from domain experts or professional rules, especially in IoT data. Though knowledge-driven methods, such as fault tree analysis (FTA), expert system (ES) \cite{Braatz2001Fault}, Dynamic Bayesian Network (DBN) \cite{DBLP:journals/sadm/FujimakiNTSY09}, have been developed for anomaly and fault diagnosis tasks, it still has limitation in knowledge modeling. Moreover, the incompleteness and fuzziness add to the difficulty in utilizing the knowledge.
\newline \indent (3). \emph{Lack of scalability in (IoT) big data}.
Since that current knowledge-based methods usually focus on small-scale specific scenarios, neither the knowledge nor the method can be easily adapted to other scenarios.

\indent Referring to the desirable properties of causality analysis in violation detection
\cite{DBLP:conf/sigmod/ChalamallaIOP14,DBLP:books/acm/IlyasC19}, we summarize that a  high-quality  explicable anomaly detection approach in industry applications always focuses on the following objectives.
\newline \indent $\bullet$ \emph{Coverage}. The solution of the method is expected to comprehensively
\emph{cover} anomaly instances existing in data.
\newline \indent $\bullet$ \emph{Conciseness}. The method needs to provide a conciseness solution rather than a redundant one, for the reason that, both time and human resources are costly in the response procedure to the anomalies. In addition, the consequence of the unexpected anomalies is unpredictable. That requires the method to provide a small-scale solution for decision making  as much as possible.
\newline \indent $\bullet$ \emph{Self-update}. The method is expected to deal with the new anomaly instances whose patterns are unknown in the knowledge base or does not occur (be detected) in the historical data.
\newline \indent $\bullet$ \emph{Less tolerance of False Negative (FN)}. Though it is difficult to entirely avoid False Negative and False Positive in practical detection tasks, it demands a high performance of the method in industrial field, especially the electricity and manufacture scenarios. FN means one fails to identify untraceable anomalies in data, which is likely to result in more serious effects, compared with FP.
\newline \indent \textbf{Contributions}. Motivated by the above,
we explore the anomaly explanation problem in multivariate time series under industrial scenarios with data and knowledge combined method in this paper. Our contributions are summarized as follows.
\newline \indent (1). We formalize the anomaly explanation problem in multivariate temporal data, and design a self-contained 3-step anomaly explanation method framework for multivariate data (see Figure \ref{frame}), according to the aforesaid four objectives. The proposed framework provides a guide to navigating the explicable anomaly detection, especially in temporal data cleaning and IoT fault diagnosis techniques.
\newline \indent (2).  We apply the 4 type of constraints proposed in \cite{DBLP:journals/debu/DasuDS16} which formalize the \emph{dependence} on attributes (columns) and entities (rows) to accurately uncover the anomalies hidden in \emph{multivariate} data in violation detection step (see Section \ref{sec3}). We devise a set-cover-based algorithm $\mathsf{AEC}$ to address the anomaly explanation, and provide concise and reliable explanation solutions covering all the anomaly representations (see Section \ref{sec4}).
\newline \indent (3). We formalize and well utilize the domain knowledge to achieve the description and the explanation of the anomalies in data. We also provide knowledge update procedures and algorithms during the iteration of detection and explanation, which allows both manual intervention and automatic update (see Section \ref{sec5}).
\newline \indent (4). We conduct a thorough experiment on real-life datasets from large-scale IoT systems. Results of the comparison experiment results verify our method provides high-quality explanation solutions of anomalies.
\section{Framework Overview}
\label{sec2}
\subsection{Problem Statement}
\label{sec2.1}
\begin{figure}[h]
\centering
\includegraphics[scale=0.20]{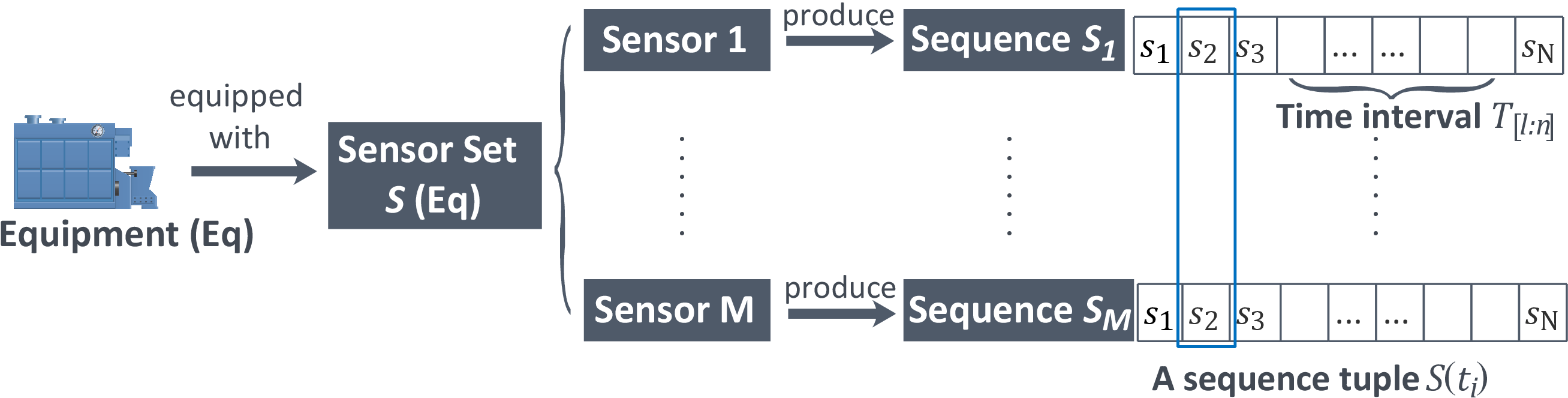}
\vspace{-1.5em}
\caption{Multivariate IoT time series.}
\label{data}
\vspace{-1em}      
\end{figure}

We outline the multivariate time series in Figure \ref{data}. $S=\langle s_1,...,s_N\rangle$ is a sequence on sensor $S$, where $N=|S|$ is the length of $S$, \emph{i.e.}, the total number of elements in $S$. $s_n= \langle x_n,t_n\rangle, (n \in[1,N])$, where $x_n$ is a real-valued number with a time point $t_n$, and for $\forall n,k \in [1,N]$, it has $(n <k) \Leftrightarrow  (t_n <t_k)$. Let $\mathsf{Eq}$ be an equipment sensor group. $\mathcal{S}_\mathsf{Eq}=\{S_1,...,S_M\} \in \mathbb{R}^{N\times M}$ is a $M$-dimensional time series, where $M$
is the total number of equipment sensors, \emph{i.e}., the number of dimensions. $T=\{t_1,...,t_n\}$ is the set of time points of time series $\mathcal{S}_{\mathsf{Eq}}$.
\newtheorem{definition}{Definition}

In this paper, we use rule-based techniques to detect the anomalies from the violation of the given constraints. We introduce the constraint set for one sequence in Definition \ref{cs}. Accordingly, given constraint $c$ for sequence $S$, $S$ is identified to \emph{violate} $c$ if the data in $S$ does not satisfy the content described by $c$. We denote such \emph{violation} by $S \not \models c$.
\begin{definition}
\label{cs}
(\textbf{Constraint set}). $C$ is the set of all constraints defined on sequence $S$, denoted by $C(S)=\{c_1,...,c_n\}$, where $c_i$ is a formulated or learnt constraint or rule the data need to meet.
\end{definition}
\begin{definition}
\label{Vfeature}
(\textbf{Violation feature}). Given sequence $S$ and the constraints set of $S$, \emph{i.e.,} $C(S)$, we maintain a 2-tuple $v = \langle S,F(c) \rangle$ of $S$ \emph{w.r.t.} constraint $c, (c\in C(S))$, where $F(c)$ is a degree function computed by a specified violation measurement $\mathcal{F}$ on $c$, which has two formats:
\newline \indent (1). If $c$ is  a qualitative constraint, \begin{equation}\label{boolean}
  F(c) = \left \{
\begin{array}{cl}
  1, & S \not \models c
  \\
  0, & S \models c.\nonumber
\end{array}
\right.
\end{equation}
\indent (2). If $c$ is  a quantitative constraint, it has $F(c) = [ d, u ]$, where $d$ and $u$ are the lower bound and the upper bound computed by the measurement $\mathcal{F}$, respectively.
\newline \indent $v$ is regarded as the violation feature of $S$ on $c$ when $S \not \models c $. $V(S) = \{v_1,v_2,...\}$ is the set of all violation features of $S$, and $\mathcal{V}(\mathcal{S})= \{V(S_1),...,V(S_M)\}$ is the total set of all violation features in sequences of data $\mathcal{S}$.
\end{definition}
It is acknowledged that the anomaly explanation discovery problem needs the assistance of knowledge provided by domain experts who have accumulated countless practical experience. The knowledge supplied from industry applications has various forms, including 1)a fault ID number which is acknowledged in the diagnostic system and can be retrieved in the users' manual, 2)some general descriptions of abnormal patterns, 3)the empirical causal inference of anomaly instances, etc. In this section, we formalize several concepts \emph{w.r.t} the provided knowledge, which is the critical input of the studied problem, besides data $\mathcal{S}$ and the constraint set $\mathcal{C}$.
\newline \indent In general, we aim to explain the detected anomalies by finding out the corresponding fault reasons. We consider an (acknowledged) anomaly event $E$ to be one reason of the occurrence of anomalies. Due to the fact that each fault event will lead to a series of unexpected changes in sensors data, we consider one change in data as one anomaly representation, denoted by $r$ in Definition \ref{r}. 
$r$ is regarded as the smallest unit of the given knowledge in our problem. We briefly present examples of a knowledge set of a sensor group in a power plant in Table \ref{Example}.
\begin{definition}
\label{r}
(\textbf{Anomaly representation}). $r = \langle S, F_r(c)\rangle$ is the anomaly representation of constraint $c$ on sequence $S$, where $F_r(c)$ is one kind of formal description depicted from domain experts or professional knowledge.
\begin{table}[t]
\centering
  \caption{Example of a knowledge set}
  \label{Example} \vspace{-0.8em}
\scalebox{0.85}{
  \begin{tabular}{|r|ll|}
    \hline
    \textbf{Event} $E$ & \textbf{Explanation} $R(E)$ &  \textbf{Representation}\\
    \hline
     Id-1 Sensor break & temperature decline,... & $\langle S_2, [-\infty, 20\%]\rangle \cdots$\\
     Id-2 Sensor break & pressure drop,... & $\langle S_1, F(c_1) \rangle$, $\langle S_3, F(c_4) \rangle \cdots$  \\
     Id-1 Engine off &  zero in power,... & $\cdots$\\
     Id-1 Boiler state instability & temperature shock,...& $\cdots$\\
   \hline
\end{tabular}}
\vspace{-1.4em}
\end{table}
\newline \indent $F_r(c)$ has the same structure of $F(c)$ referring to Definition \ref{Vfeature}. For quantitative constraints, $F_r(c) = [ d_r, u_r ]$, where $d_r$ (resp. $u_r$) is the lower (resp. upper) value of the description of the knowledge.
\end{definition}
Accordingly to Definition \ref{r}, the unexpected changes caused by an event $E$ is formally presented as a set of anomaly representations. Such a set of representations is considered as an \emph{explanation} of anomaly instances in data. That is, $R(E) = \{r_1,...,r_n\}$ is a set of anomaly representation describing one event $E$. $R(E)$ is the maximum set of representation $r$s which can be provided by domain experts. The set of all explanations, denoted by $\mathcal{R}=\{R(E_1),...,R(E_N)\}$ is the formal description of the domain knowledge provided for the equipment sensor group $\mathcal{S}_{\mathsf{Eq}}$. Formally, the problem studied in this paper is stated in Definition \ref{Problem}.
\begin{definition}
\label{Problem}
(\textbf{Problem description}). Given the multivariate time series $\mathcal{S}$ of equipment $\mathsf{Eq}$, the constraint set $\mathcal{C}$, and the knowledge set $\mathcal{R}$, the anomaly explanation problem includes two tasks below, with the four objectives proposed in Section \ref{sec1}: (1) to detect the violations in $\mathcal{S}$ according to $\mathcal{C}$, locate the violated sequence ID with time interval $T$, and record the violations in the set $\mathcal{V}(\mathcal{S})$, and (2) to discover an explanation set $\mathcal{R}' \subseteq \mathcal{R}$ \emph{w.r.t} $\mathcal{V}(\mathcal{S})$.
\end{definition}
\subsection{Overview of the Approach}
\label{sec2.2}
\begin{figure*}[t]
\centering
\includegraphics[scale=0.2]{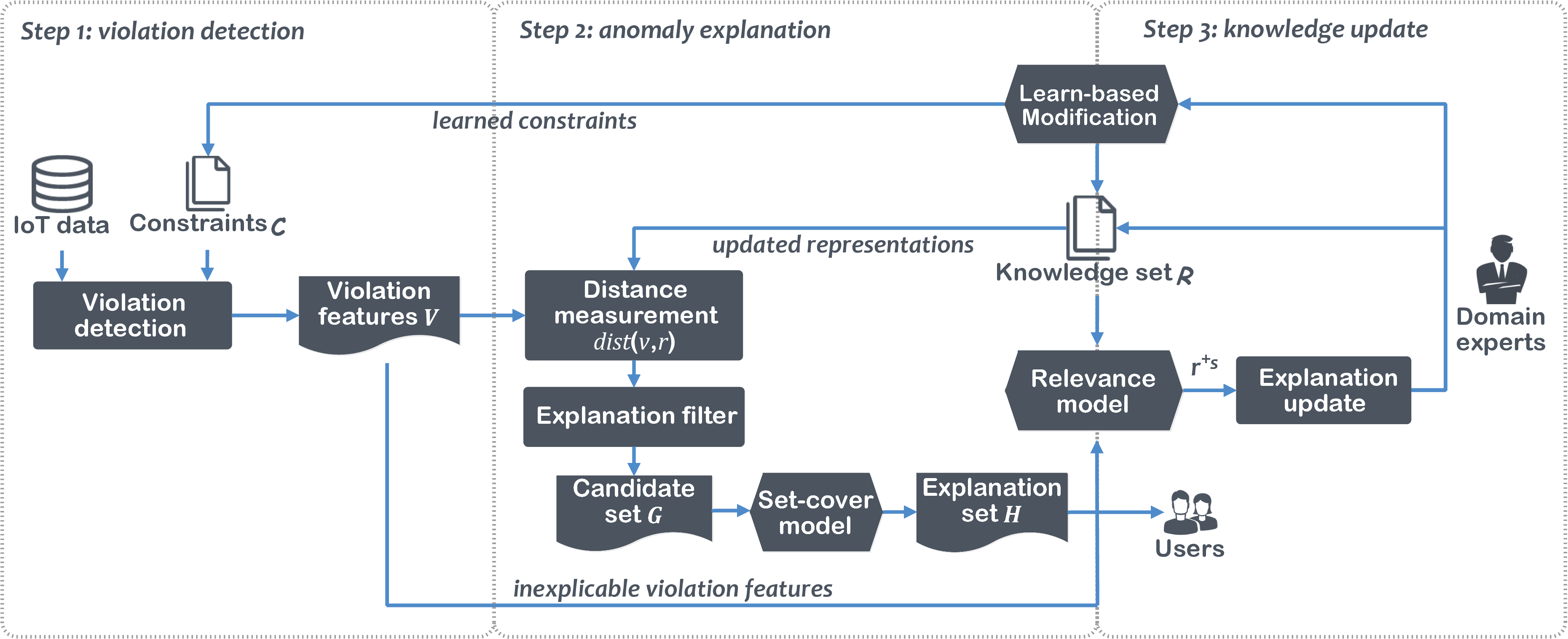}
\vspace{-0.8em}
\caption{Method framework overview}
\label{frame}
\vspace{-1.5em}      
\end{figure*}
Figure \ref{frame} outlines our method, which contains three phases: violation detection, anomaly explanation, and knowledge update. We will discuss the violation detection with types of constraints in temporal data in Section \ref{sec3}, and introduce our anomaly explanation algorithms in Section \ref{sec4} in detail. The knowledge update step will be discussed in Section \ref{sec5} with the procedure of Algorithm \ref{update} and the function in Algorithm \ref{AEUP} to find out the candidate update explanations.
\section{Discovery of anomaly instances}
\label{sec3}
\subsection{Constraint-based anomaly detection}
\label{sec3.1}
Since that the dependance and relevance does exists among \emph{multivariate} time series, we apply 4 types of constraints discussed in \cite{DBLP:journals/debu/DasuDS16} in our violation detection process. As shown in Table \ref{Types}, the 4-Type constraints embody the dependence on attributes (columns) and entities (rows) for temporal data.
\begin{table}[h]
\centering
  \caption{Types of constraints}
  \label{Types}\vspace{-0.8em}
  \begin{tabular}{|c|cc|}
    \hline
    \textbf{Type} &\textbf{Singe column} &\textbf{Multi-column}\\
    \hline
    \textbf{Single row }&  Type 1 & Type 2\\
    \textbf{Multi-row} & Type 3& Type 4\\
   \hline
\end{tabular}
\end{table}
\newline \indent Accordingly, we summarize some instances of the four-type constraints in Figure \ref{freq}. We consider the value domain of data points in sequence as the simple instance of Type-1 constraints, CFD for relational data and Physical Mechanism for industrial data are concluded as multi-sequence constraints. Constraints, such as SD, SC, and VC, formalizing the dependence of data points along the time in one sequence belongs to Type-3 constraints. Rules describing the similarity located in multi sequences can be classified as Type-4 constraints. Various types of constraints assist to precisely locate the anomalies and it has potential to uncover the anomalies as early as possible.
\begin{table}
\centering
  \caption{Examples of constraint types}
  \label{freq} \vspace{-0.8em}
\scalebox{0.9}{
  \begin{tabular}{|c|ll|}
    \hline
    \textbf{Type} &\textbf{Singe sequence} &\textbf{Multi-sequence}\\
     \hline
    \textbf{Single time point} &  T-1: Value domain & T-2: CFDs\\
     &  from documents & T-2:  Physical Mechanism\\
     \hline
    \textbf{Time interval} & T-3: SD, SC\cite{DBLP:conf/sigmod/SongZWY15}, & T-4: Similarity Constraints\\
     & T-3: Variance Constraints\cite{DBLP:conf/dasfaa/YinYWHL18}&  \\
  \hline
\end{tabular}}
\vspace{-1.4em}
\end{table}
\vspace{-1em}
\subsection{Anomaly distance measurement}
\label{sec3.2}
With types of constraints, we can detect the violations hidden in sequences. After we obtain violation features, we need to compare them with the given anomaly representations to determine the real-happened anomaly events. We first propose the  concept of ``\emph{Explicable}'' in Definition \ref{Explicable feature}, and propose anomaly distance function in Definition \ref{distance}, according to which we are able to quantize the closeness of violations with acknowledged anomaly events.
\begin{definition}
\label{Explicable feature}
(\textbf{Explicable feature}).
Given a  detected violation feature $v: \langle S, F(c)\rangle$, $v$ is explicable  by $\mathcal{R}$, iff $\exists \, r:\langle S', F_r(c')\rangle\in R, R\subseteq  \mathcal{R}$, $S=S'$ and  $c=c'$.
\end{definition}
\begin{definition}
\label{distance}
(\textbf{Anomaly distance}). Given feature $v: \langle S, F(c)\rangle$ and the (corresponding) representation $r: \langle S, F_r(c)\rangle$,
the distance between $v$ and $r$ \emph{w.r.t} constraint $c$ is computed as follows,
\newline \indent If $c$ is a quantitative constraint, then it has
\begin{equation}
\label{e1}
dist\big(v,r\big) = dist\big(F(c),F_r(c)\big)= 1-\frac{[d,u]\cap[d_r,u_r]}{[d,u]\cup[d_r,u_r]},
\end{equation}
\newline \indent If $c$ is a qualitative constraint, then it has
\begin{equation}
\label{e2}
dist\big(v,r\big) = dist\big(F(c),F_r(c)\big)=|F_r(c)-F(c)|.
\end{equation}
\end{definition}
The proposed distance function $dist(\cdot,\cdot)$ is only measurable with regard to the \emph{same} constraint. It does not make sense to compute the distance between $v$ and $r$ which hold different constraints.
Obviously, the anomaly distance $dist(v,r)$ coincides with the properties of the \emph{distance} function, which locates in $[0,1]$ and the value of $dist(v,r)$ is lower as feature $v$ is closer to representation $r$. More specifically, for qualitative constraints, $dist(v,r)$ only has two values where $dist(v,r) = 0$ shows the detected $v$ is consistent with the representation $r$, and $dist(v,r) = 1$, otherwise. For quantitative constraints, $dist(v,r) \in (0,1)$ shows $v$ is partially \emph{consistent with} $r$, while $dist(v,r) = 1$ indicates that feature $v$ is completely different from the representation $r$, with $[d,u] \cap [d_r, u_r] = 0$. Thus, we describe how to determine whether the feature is consistent with the representation in Definition \ref{consistent}.
\begin{definition}
\label{consistent}
Given feature $v$ and the corresponding representation $r$,
$v$ is identified to be \textbf{consistent with} $r$ if having
\begin{equation}
\label{econsistent}
  dist(v,r)  \left \{
\begin{array}{cl}
  = 0, & \rm{c \, is \, qualitative},
  \\
  < 1, & \rm{c \, is \, quantitative}. \\
\end{array}
\right.
\end{equation}
\end{definition}
More generally, $dist(v,r) < 1$ is too loose to estimate whether the feature matches a representation. Thus, we introduce a threshold $\theta\in(0,1)$ and consider $v$ is consistent with $r$ when $dist(v,r) < \theta$. $\theta$ can be learned from amount of experiments, or set manually as required.
\newline \indent We highlight that, one may fail to find the one-to-one match between each violation feature and each representation in the real industry scenarios. That is, a violation feature $v$ obtained from the violation detection method is not necessarily identified to be explicable (by $\mathcal{R}$). It results from both the precision limitation of violation detection techniques and the incompleteness in knowledge representations provided by experts. In the next section, we will introduce our anomaly explanation approach, in which we take into consideration the multiple conditions between the detected features and the given representations in detail.
\section{Identifying Anomaly Explanations}
\label{sec4}
\subsection{Candidate Explanations Discovery}
\label{sec4.1}
In anomaly explanation and analysis research, especially the knowledge-based study, it is acknowledged that both \emph{incompleteness} and \emph{ambiguity} always exist in experts knowledge. For the former, domain experts may not provide all descriptions about one anomaly instance reflected in the sensor data. The main reasons include 1) experts' limitations in professional degree and human's understandability of anomaly problems, and 2) the deficient deployed sensors, which may fail to record some parameters that are critical to the explanation of abnormal data.
For the latter, since some knowledge are accumulated form practical experience, one cannot expect industry experts to be always definite about they explanations. Part of these explanations may happen with probability.
\newline \indent According to our investigation of manufacturing and the electricity industry, both definite and presumable knowledge are applied in  anomaly explanation and fault diagnose tasks. In this case, we divide the explanations of anomaly into two categories: \emph{exact explanations}  and \emph{possible explanations}.
\newline \indent
Given a fault event $E$, the \textbf{exact explanation} of $E$ is the maximum set of anomaly representations, denoted as $R^*(E)$, in which
the fault event $E$ leads to the existing of representations $\{r_1,...,r_x \}$ in data. $\forall i\in[1,x], r_i$  is called an exact representation for $E$. The \textbf{possible explanation} of $E$ is a set of anomaly representations, denoted by $R^+(E) = \{r_1,...,r_y \}$, in which $E$ lead to the existing of anomaly representation $r_i, i\in[1,y]$ with probability $\Pr, \big(\Pr\in (0,1)\big) $. $\forall j\in[1,y], r_j$ is called a possible representation for $E$.

With the two categories, we denote the explanation $R$ by the combination of $R^*$ and $R^+$, as shown in Proposition \ref{proposition1}.
\newtheorem{proposition}{Proposition}
\begin{proposition}
\label{proposition1}
The explanation $R$ \emph{w.r.t.} event $E$ is the union of the exact explanation and the possible explanation of $E$, denoted by $R=R^* \cup R^+$, where $R^* \cap R^+ = \emptyset$ and $R^* \neq\emptyset$.
\end{proposition}
In general, the exact  $R^*$ is considered as the key factor in the identification of $E$, while the possible $R^+$ helps to describe the event  in a rough way. In industry scenarios, the occurrence of a fault, especially a known one, will \emph{certainly} give rise to the violation of a series of constraints, as described by $R^*$. But on the contrary, one cannot be sure that whether the event really happens when some representations in $R^*$ have been detected. We formally describe the relationship between a fault $E$ and its explanation $R^*(E)$ in Proposition \ref{proposition2}. Accordingly, we are able to make further analysis in order to obtain the fault set and provide a reliable and high-quality explanation of the anomalies.
\begin{proposition}
\label{proposition2}
Given a fault event $E$ with $R(E)$, $E$ is the sufficient and unnecessary condition of its exact explanation $R^*(E)$, $(R^*(E) \subseteq R(E))$.
\end{proposition}
From the above, we are able to narrow the computation on $\mathcal{R}$ by finding out a subset of $\mathcal{R}$, in which all exact representations have appeared in data. Such subset of $\mathcal{R}$ is denoted as $G = \{R_1,R_2,... \}$  as shown in Definition \ref{G}. Thus, we first find out the candidate explanation set $G$ from $\mathcal{R}$ by \{verifying the appearance of exact explanations with all violation features\} detected by the previous steps, and then precisely compute the explanation set from the candidate result $G$.
\begin{definition}
\label{G}
Given the violation feature set $\mathcal{V}$ of  $\mathcal{S}$ \emph{w.r.t.} $T$, and the set $\mathcal{R}$, $G$ is identified as a \textbf{candidate explanation set} if satisfying: 1) $G\subseteq \mathcal{R}$, and 2) $\forall R \in G, R^* \subseteq R$, $\forall r:\langle S,F_r(c) \rangle \in R^*$, $\exists \, v \in \mathcal{V}$, $v$ is consistent with $r$ \emph{w.r.t.} $c$.
\end{definition}
The candidate explanation set discovery process is shown in Algorithm \ref{findG}. After initialize an empty set $G$, we enumerate each explanation set $R$ from $\mathcal{R}$ in the outer loop (Lines 2-9), and maintain a label \emph{cand} to record whether all  elements in $R$ exist in the detected violation features. Within the outer loop,  we enumerate each representation $r$ from the exact explanation set of $R$, \emph{i.e.,} $R.R^*$ and identify the occurrence
of $r$ in data. We let the label $\emph{cand} = 0$ when the sequence $S$ does not violate constraint $c$ \emph{w.r.t} $r$, or
the violation in $S$ is different from $r$, \emph{i.e.,} $dist(v,r) = 1$ (Lines 5-7). After all exact representations in $R.R^*$ are visited, we add all $R$s with  label $\emph{cand} = 1$  into $G$, and finally obtain the objective subset $G$ from $\mathcal{R}$.
\begin{algorithm}[t]
\label{findG}
\begin{small}
\caption{Compute Candidate Explanations}
\LinesNumbered
\KwIn{$\mathcal{V}=\{V(S_1),...,V(S_M)\}$: the set of violation features  in $\mathcal{S}$ \emph{w.r.t} $T$,  the explanation set $\mathcal{R}$ }
\KwOut{the set of candidate explanations $G$}
initialize $G \leftarrow \emptyset$\;
\ForEach {$R\in \mathcal{R}$}
{
$\emph{cand} \leftarrow 1$\;
\ForEach{$r \in R.R^*$}{
\If{ $v.F(c) = 0 $ or $dist(v,r) = 1$ }{
$\emph{cand} \leftarrow 0$\;
break\;
}}
\If{cand $= 1$}{
$G \leftarrow G \cup\{ R \}$\;
}
}
\Return $G$\;
\end{small}
\end{algorithm}
\vspace{-0.5em}
\subsection{Cost-based Explanation determination}
\label{sec4.2}
According to Algorithm 1, we can \emph{qualitatively} find out the candidate set which contains an overall explanation of fault that really happens in data. However, the set $G$ is far from a high-quality solution for industry applications.
In the next step, we aim to make further discovery of anomaly explanations considering the four objectives as discussed in Section \ref{sec1}.
\newline \indent Note again that an explanation includes a few representations, either the exact or the possible ones. We identify the explanations of anomalies by measuring the distance between the violation feature and the representation \emph{w.r.t.} the same $c$. Intuitively, we should choose those explanations with close distance values between the detected features and the given representations. In order to model the distance degree between $v$s and $r$s, we introduce a cost-based principle to quantitatively compute how much the violation features matches an anomaly explanation in Definition \ref{cost}.
\begin{definition}
\label{cost}
(\textbf{Explanation Cost}). The cost of applying event $E$ as an explanation of the anomalies in $\mathcal{S}$ is,
\begin{equation}
\label{equationcost}
Cost(E) = \frac{\sum_{r_i \in R^*}^x  dist(v_i,r_i)}{|r_i.S|} + \frac{\sum_{r_j \in R^+}^y  w \cdot dist(v_j,r_j)}{|r_j.S|}, \nonumber
\end{equation}
where $R^*(E)$ (resp. $R^+(E)$) is the exact (resp. possible) explanation of $E$, $v$ is the detected violation feature of sequence $S$ \emph{w.r.t} constraint $c$, $dist(v,r)$ is the anomaly distance between $v$ and $r$ \emph{w.r.t}  $c$, and $|r.S|$ is the number of sequences involved in $r$, and $\omega \in (0,1)$ is the probability value of a possible representation in $R^+$.
\end{definition}
According to Definition \ref{cost}, a higher $Cost(E)$ value shows that the violation features in data less match the fault representations of $E$, and intuitively, fault $E$ is less likely to be a reason of the detected anomalies. It is acknowledged that there probably exists more than one fault in the equipment system at the same time interval, which urges us to explore the multiple reasons which can \emph{cover} all the anomalies detected by constraints. Here, we introduce \emph{the minimum weight set covering problem (MWSCP)} in our method to derive the optimal explanations of anomalies.
\newline \indent Note that we are faced with
different cases in the comparison of the observed (\emph{a.k.a} detected) real data and the knowledge representations. The constraint-based anomaly detection phase identifies the input data as two categories: 1) abnormal data which violate at least one constraint, and 2) normal data which do not have violations. Considering the given knowledge, there also has two cases: 1) existing representation which describes the violation of certain constraint, and 2) no given representations at all.
Accordingly, there are totally four kinds of conditions to be considered in the explanation analysis phase, as shown in Table \ref{ABCD}. From the point of the monitoring data, we summarize the total four kinds of data as Set A, B, C, D, respectively.
\begin{table}
\centering
  \caption{Confusion matrix in the explanation phase}
  \label{ABCD} \vspace{-0.8em}
  \begin{tabular}{r|ll}
    \hline
  \textbf{Knowledge} \,\,\,\,\,\,&  \multicolumn{2}{c}{\textbf{Detected}}
    \\ \cline{2-3}
     & \textbf{Abnormal data}  & \textbf{Normal data} \\
   \hline
     \textbf{Exist representations}  & Set A: $\mathcal{V}^*$ & Set B   ($\mathcal{R} \setminus\mathcal{R}^o$)\\
  \textbf{No  representation} & Set C: $\mathcal{V}\setminus\mathcal{V}^*$ & Set D\\
 \hline
\end{tabular}
\vspace{-1.5em}
\end{table}
\newline \indent Set D in Table \ref{ABCD} is beyond our research, for data in Set D is neither detected to be abnormal nor described by knowledge. We focus on the analysis of Set A, B, and C. Specifically, Set A contains the violated data which can be explicable by representations from knowledge set $\mathcal{R}$. Set B contains the data which have representations in $\mathcal{R}$ but are identified as normal. Set C contains the data detected to have violations while there are no representations in $\mathcal{R}$ to explain these violations. It mainly has two reasons: \emph{i}) The constraint instances are set too strict so that the method falsely identifies some normal data to be abnormal, or \emph{ii}) some new anomaly patterns are discovered which is unknown in the present knowledge set. Thus,
the cases in Set C are also serious in our solutions, because it can assist update the knowledge set.
\newline\indent Below we first introduce how to precisely explain Set A, and we will propose the updating method according to the detected Set C in the next Section. Considering our four objectives, our solution needs to cover all the detected anomalies with a small-size results. That is,  to find out a concise set of explanations from $\mathcal{R}$ which could explain all the violations. Moreover, our method is expected to provide explanations much well describing the anomalies with close distance values.
\vspace{-0.2em}
\subsection{Set-Cover-based anomaly explanation}
\label{4.3}
As discussed above, we aim to find a subset of $\mathcal{R}$ as the problem solution which \emph{covers} all the violated instances in data. We denote the data in Set A as $\mathcal{V}^*=\{v:\langle S, F(c) \rangle | v\in \mathcal{V}(\mathcal{S})$, and $v$ is explicable by $\mathcal{R}\}$. We apply \emph{the minimum weight set covering problem (MWSCP)} \cite{DBLP:books/daglib/p/Karp10},\cite{1992Enhancing} to solve our AE problem. Definition \ref{coverproblem} formalizes our problem, where the set $\mathcal{V}^*$ is the target to be covered, and the explanation cost $Cost(R)$ is regarded as the weight.
\begin{definition}
\label{coverproblem}
(\textbf{Anomaly Explanation Problem}) Given the violation set $\mathcal{V}(\mathcal{S})$ \emph{w.r.t}. $\mathcal{S}$, the knowledge set $\mathcal{R}$, and the candidate explanation set $G$. Our anomaly explanation problem is to find an explanation set $H$ which satisfies
\begin{alignat}{2}
\min\quad & \sum_{R(E)\in H} cost(E) &{}&
\nonumber \\
\mbox{s.t.}\quad
&\quad H \subseteq G, &{}&\nonumber \\
&\bigcup_{R^o(E) \subseteq R(E)} R^o.cover = \mathcal{V}^*, &\quad& \forall R(E) \in H \nonumber
\end{alignat}
where $R^o(E) =\{r_1,...r_m\}$ is a subset of $R(E)$ \emph{w.r.t} anomaly event $E$ having $\forall \, r_i \in R^o(E), i\in[1,m], \exists$ explicable feature $v\in \mathcal{V}^*, dist(v,r_i) < 1$, and 
$R^o.cover$ represents the total set of the violation features which are consist with $r$s in $R^o$.
\end{definition}
\indent \indent \textbf{The greedy-based heuristic algorithm}. Since the set cover computing is NP-hard \cite{DBLP:books/acm/IlyasC19,1992Enhancing}, we introduce greedy-based algorithms for our AE problem.
Considering the coverage issues, the solution  of our anomaly explanation problem proposed above is a covering of the set $\mathcal{V}^*$ first, and then it satisfies the minimum cost principle. In this case, we are able to find out whether an explanation $R \in \mathcal{R}$ is certainly contained in the
the solution $H$ or certainly does not exist in $H$.
Both cases are concluded in Proposition \ref{proposition3} and Proposition \ref{proposition4}, respectively. Proposition \ref{proposition3} shows that if the explanation $R_j$ covers more violations besides $R_i.cover$, while $R_j$ has little cost than $R_i$, $R_i$ will not be selected into the solution $H$. Proposition \ref{proposition4} shows that an explanation $R$ must be selected into the solution if it is the only explanation that can covers violation $v$.
\begin{proposition}
\label{proposition3}
Given $\mathcal{R}$, the explanation $R_i$ is not valid and
does not exist in the solution $H$, if $\exists \, R_i, R_j \in \mathcal{R}, R_i.cover \subseteq R_j.cover$, and $Cost(R_i) > Cost(R_j)$.
\end{proposition}
\begin{proposition}
\label{proposition4}
$R$ from $\mathcal{R}$ is a valid explanation and does exist in the solution $H$, if $\exists \, v\in \mathcal{V}^*$, there is only one explanation $R$ in $\mathcal{R}$ which satisfies $v \in R.cover$.
\end{proposition}
Considering both effectiveness and efficiency, we propose a greedy-based heuristic algorithm to obtain $H$. The general principle is, to give priorities to choosing the explanation \emph{i}) which has smaller cost value, and \emph{ii}) covering the violations of constraints defined on multiple sequences, specifically the Physical Mechanism constraints in this paper. For the former, it is obvious that the fault event with smaller cost value is more reliable to explain part of the anomalies. For the later, we consider to cover the violations existing in multiple sequences prior to the ones in single sequence, for three main reasons: 1) the violation between sequences are more likely to involve fault event(s) than the violation happening in single sequence. Because the detected single-sequence violations may just occur sporadically for some reasons, which does not require an explanation. The faults which happen in several sensors are always more serious than the ones only happen in one sensor,
and 2) the multi-sequences violations always contains much more features than single-sequence violations. The process of multi-sequences violations contributes to increasing the coverage of the solution.
\begin{algorithm}[t]
\label{findH}
\begin{small}
\caption{Compute Explanation Set}
\LinesNumbered
\KwIn{the set $\mathcal{V}^*$ of violation features  in $\mathcal{S}$ \emph{w.r.t} $T$,  the candidate explanation set $G$ }
\KwOut{the set of final explanations $H$}
initialize $H \leftarrow \emptyset$\;
delete $R$s from $G$ according to Proposition \ref{proposition3}\;
insert $R'$s from $G$ into $H$ according to Proposition \ref{proposition4}\;
$G \leftarrow G \setminus R's$\;
$\mathcal{V}^* \leftarrow \mathcal{V}^* \setminus R'.cover$\;
select the subset $\mathcal{V}_{M}^*$ from $\mathcal{V}^*$ where $\mathcal{V}_{M}^*=\{v| v.K > 1$ and $v\in \mathcal{V}^*\}$\;
sort all $v$s in $\mathcal{V}_{M}^*$ in the descending order of the size of $v.K$\;
\ForEach{$v \in \mathcal{V}_{M}^*$}{
$R(E) \leftarrow \arg \min_{v\in R(E).cover} Cost(E)$\;
$H \leftarrow H \cup \{ R(E)\}$\;
$\mathcal{V}_{un}^* \leftarrow \mathcal{V}^* \setminus H.cover$\;
}
\While{$\mathcal{V}^* \ne \emptyset$}{
$R(E) \leftarrow \arg\max_{R(E)} \frac{|R.cover \cap \mathcal{V}_{un}^*|}{Cost(E)}$\;
$H \leftarrow H \cup \{ R(E)\}$\;
$\mathcal{V}_{un}^* \leftarrow \mathcal{V}_{un}^* \setminus R.cover$\;
}
\Return $H$\;
\end{small}
\end{algorithm}
\newline \indent Algorithm \ref{findH} outlines our heuristic algorithm, which mainly consists of three steps, as discussed below.
\newline \indent \emph{Global optimization} (Lines 1-5). After initializing an empty set $H$, we first execute the global optimization according to  Proposition \ref{proposition3} and Proposition \ref{proposition4}. Thus, we narrow the size of the input $G$ by deleting the invalid explanations $R$s, while we insert the valid explanations $R$'s into set $H$. After that, we delete $R$'s from $G$ and correspondingly delete $R'.cover$ from the violation set $\mathcal{V}^*$.
\newline \indent \emph{Covering multi-sequences violations} (Lines 6-11). After we deal with all the valid and invalid explanations, we begin to select explanations from the present set $G$ to cover the multi-sequences violations. We sort all multi-sequence features in the  descending order of the number of sequences involved in each feature $v$. We  then enumerate each feature $v$ from the sorted set $\mathcal{V}_M^*$, and greedily find an fault $E$ whose explanation $R(E)$ which can cover $v$ with the minimum $Cost(E)$ value. We put such $R$ into the solution set $H$, and
finish the iteration when all the features in $\mathcal{V}_M^*$ have been visited. We then delete all the violation features covered by $H$ from $\mathcal{V}^*$ and let the set be $\mathcal{V}_{un}^*$, which needs to be covered in the following step.
\newline \indent \emph{Covering single-sequence violations} (Lines 12-15).
Faced with $\mathcal{V}_{un}^*$, we compute the total number of $v$s in $\mathcal{V}_{un}^*$ covered by the same explanation $R$, denoted by $|R.cover \cap \mathcal{V}_{un}^*|$, and we iteratively choose the explanation $R$ which has the maximum ratio  of the above number to  $Cost(E)$. Correspondingly, we add $R$ into $H$ and then delete $R.cover$ from the present set $\mathcal{V}_{un}^*$. This iteration finishes until there is no features in $\mathcal{V}_{un}^*$, and we finally obtain the solution $H$.
\newline \indent \emph{Complexity}. The modification process in Algorithm \ref{findH} lines 2-5 costs $O(|\mathcal{V}|\cdot |\mathcal{R}|^2)$ time, and the sorting in line 6 costs $O(|\mathcal{V}|\cdot \log |\mathcal{V}|)$. Generally, the loops Lines 8-11 and Lines 12-15 both
 cost $O(|\mathcal{V}|^2 \cdot |\mathcal{R}|)$ at worst. To put it together, Algorithm \ref{findH} spends $O(|\mathcal{V}| \cdot |\mathcal{R}|\cdot \max\{|\mathcal{V}|,|\mathcal{R}|\})$.
\section{Knowledge Update}
\label{sec5}
Though we find out a solution of explaining the detected anomalies with the \emph{existing} reasons in Algorithm \ref{findH}, there remains some anomalies which are inexplicable by the knowledge set, \emph{i.e.}, the Set C in Table  \ref{ABCD}. There are mainly several reasons of the occurrence of the violated data in Set C: (1) Some of the explanations \emph{w.r.t} a fault are not reliable enough to conclude the corresponding fault event. That is, the representations in such explanation are far from precise which fails to identified the fault from the violation features. (2) New fault events are discovered by the constraint set $\mathcal{C}$ which are not known in the present knowledge set $\mathcal{R}$. With the both cases, we aim to update and improve the present knowledge set according to the detected results. The updated knowledge set will provide more precise anomaly explanations in return.
\newline \indent In this section, we propose our update strategies faced with the inexplicable violations. We first discuss the update of anomaly representations, especially the possible representations, utilizing the relevance between the detected violation features in Section \ref{sec5.1}, and we introduce a knowledge set modification strategy in Section \ref{sec5.2}, which assists improve the quality of the knowledge set from the iterations of anomaly explanation and knowledge update.
\subsection{Update of Anomaly Representations}
\label{sec5.1}
As discussed above, the imperfect description concluded in $\mathcal{R}$ is one of the most serious reasons which results in the remaining inexplicable anomalies. Faced with Set C in Table \ref{ABCD}, we consider to update the explanation of fault events by either adding new representations to an explanation or directly adding an explanation \emph{w.r.t}. a new fault event. We first consider to update the existing representations within a fault's explanation with a relevance analysis of the violation features, and then consider to create a new record of the unknown anomalies in $\mathcal{R}$.
\newline \indent 
When we try to find the faults to explain the anomalies, there probably remains some violation instances that we fail to find the fault reasons to cover them. We need to update and modify the representations of faults in order to improve the description of faults.  Faced with the update task, we generally consider to insert some new violation features to the explanation of a fault, where these new features are regarded as supplementary (possible) representations of the existing explanation. Thus, we are able to find a more precise solution $H'$ to cover the detected violations.
\newline \indent When real faults happen, the violations \emph{w.r.t} one fault probably do not occur individually. One one hand, it is possible that the anomaly in one sequence $S$ brings about the multiple violations of different constraints on $S$. On the other hand, some violations \emph{w.r.t} a multi-sequence constraint $c$ would occur at the same time in the involved sequences, \emph{i.e.}, $c.domain$.
To achieve the interpretability and the dependability of the representation update, we introduce the \emph{relevance} analysis between the existing knowledge and the learnt violation features. We discuss the \emph{relevance} in Section \ref{sec5.1.1}, and then propose our update algorithm in Section \ref{sec5.1.2}.
\subsubsection{Relevance in anomaly representations}
\label{sec5.1.1}
Suppose $r^+_1$ and $r^+_2$ are two anomaly representations learnt from violation features after  lots of iterations of detection and update. The relevance between them is formalized in Definition \ref{drelevance}. We consider the different representations in the same sequence to be directly related to each other. Besides, the representations in different sequences come from one multi-sequences constraints are also directly related to each other.
\begin{definition}
\label{drelevance}
(\textbf{Relevance between $r$s}). Given two anomaly representations $r_1: \langle S_i, F(c_m) \rangle $ and $r_2: \langle S_j, F(c_n) \rangle$, $r_1$ is \emph{related to} $r_2$, denoted by $r_1 \leftrightarrow r_2$, if it satisfies either
\newline \indent (1). $S_i$ and $S_j$ are the same sequence, while $c_m$ and $c_n$ are different constraints, \emph{i.e.}, $i = j$, and $m \neq n$, or
\newline \indent (2). $S_i$ and $S_j$ are different sequences, while $c_m$ and $c_n$ are the same constraint. Specifically, $c_m$ (\emph{a.k.a} $c_n$) is a multi-sequence constraint whose domain contains $S_i$ and $S_j$.
\newline \indent Otherwise, $r_1$ and $r_2$ are not related to each other, denoted by $r_1 \nleftrightarrow r_2$.
\end{definition}
The relevance between $r$s is a symmetrical relation, while it does not has transitivity because there are two factors in the identification of the relation ``$\leftrightarrow$'', \emph{i.e,} $r$s \emph{w.r.t} the same sequences, and $r$s \emph{w.r.t} the same constraints.
With the relation ``$\leftrightarrow$'', we are able to update $R$s according to the relevance between a learnt representation $r^+$ and an existing representation in $R$. Such learnt $r^+$s come from the detected violation features, as formalized in Proposition \ref{CoverAE}.
\begin{proposition}
\label{CoverAE}
Given feature $v$, let $\emph{CoverAE}(v)=\{R| R\in \mathcal{R}, v\in R.cover\}$ be the set of all explanations which cover the occurrence of $v$. Feature $v^*$ is identified to be a learnt representation of the anomaly event $E$ described by $R$, if it satisfies (1) $v^*$ is uncovered by the solution $H$, and (2) $v^*$ is related to $v$.
\end{proposition}
\subsubsection{Update Algorithm}
\label{sec5.1.2}
We propose the update process of the knowledge set $R$ in Algorithm \ref{update}. We first construct a graph $\mathcal{G}$ where each vertex denotes a violation feature, and there exists an edge between two vertices $v_i$ and $v_j$ if they are  related to each other according to Definition \ref{drelevance}. We maintain a set of explanations $\emph{CountAE}$ for each feature $v$, in which each element (\emph{i.e.}, explanation) is able to cover $v$ (Lines 2-3). After initializing the visit flag of features, we begin to iteratively visit uncovered features and update them into the existing explanations or create a new anomaly explanation (Lines 5-13). Concretely, for an uncovered feature $v$, we compute all the features which are related to $v$ (denoted by $\mathsf{Candr}$), and obtain all the explanations which should be updated \emph{w.r.t} $v$ (denoted by $\emph{UpSet}$) in Line 6. This process is implemented by a function $\textsc{FindUp}$ as discussed in Algorithm \ref{AEUP} below. If the set $\emph{UpSet}$ is empty, which means there does not exist any anomaly events that can potentially explain the occurrence of $v$, we consider to create a new anomaly event described by $v$ and its related features (Lines 7-8). We enumerate each violation feature $v$ in set $\mathsf{Candr}$, update the set $\emph{CountAE}(v)$ with the new explanation, and insert $v$ into $R_{\mathsf{up}}$ as a new representation $r^+ = \langle S, F(c) \rangle$ with a initial weight $w$ (Lines 10-12). After finishing the update process of $\mathsf{Candr}$, we delete all the elements in $\mathsf{Candr}$ from $\mathcal{V}_{\mathsf{uncover}}$, and continue to process next feature in $\mathcal{V}_{\mathsf{uncover}}$. We obtain the undated knowledge set $\mathcal{R}$ until all uncovered features have been visited.
\newline \indent We then introduce the proposed function \textsc{FindUp} in Algorithm \ref{AEUP}, where we achieve to find out (1) all uncovered features related to the input feature $v$, denoted by the set $\mathsf{Candr}$, and (2) all the anomaly explanations which need to be updated \emph{w.r.t} $v$, denoted by \emph{UpSet}. Given a feature $v$, we first mark that $v$ has been visited, and then we determine whether $v$ has been already covered by acknowledged explanations. If so, $v$ will not be considered as a candidate new representation. $\emptyset$ and
the present set $\emph{CountAE}$ will be returned (Lines 2-3). Otherwise, when the present feature is not covered by any anomaly events, we initialize the set $\mathsf{Candr}$ with $v$ and the set \emph{UpSet} with $\emptyset$, and begin to add elements into the both set. We iteratively visit the uncovered features related to $v$, and complete the set $\mathsf{Candr}$ as well as find out all the existing explanations to be updated \emph{w.r.t} $v$. After the loop in lines 5-9 finishes, both set $\mathsf{Candr}$ and \emph{UpSet} will be returned to Algorithm \ref{update}.

 \begin{algorithm}[t]
\label{update}
\begin{small}
\caption{Explanation Update}
\LinesNumbered
\KwIn{the knowledge set $\mathcal{R}$, the set $\mathcal{V}$ of all detected violations, the set $\mathcal{V}_{\mathsf{uncover}}$ of violation features uncovered by $H$}
\KwOut{the updated knowledge set $\mathcal{R}$}
construct graph $\mathcal{G}=(\mathcal{V},\mathcal{E})$ where each identified feature $v$ denotes a vertex in $\mathcal{G}$, and the
edge $(v_i,v_j) \in E$ exists if having $v_i$ related to $v_j$\;
\ForEach{$v \in \mathcal{V}$}{
$\emph{CoverAE}(v) \leftarrow \{R|R \in \mathcal{R}, v \in R.{cover}\}$\;
}
Initialize $\forall \, v\in \mathcal{V}$, $\emph{Flag}[v] \leftarrow \textrm{False}$\;
\ForEach{$v \in \mathcal{V}_{\mathsf{uncover}}$}{
$\mathsf{Candr}$, \emph{UpSet} $\leftarrow \textsc{FindUp}(v)$\;
\If{UpSet $= \emptyset$}{
create a new anomaly event and update $R_{\mathsf{up}}$ into \emph{UpSet}\;
}
\ForEach{$v \in  \mathsf{Candr}$}{
$\emph{CoverAE}(v) \leftarrow \emph{UpSet}$\;
\ForEach{$R_{\mathsf{up}} \in$ UpSet}{
insert $v$ into $R_{\mathsf{up}}$ as a new possible representation $r^+$ with  the initial weight $w$\;
}
}
$\mathcal{V}_{\mathsf{uncover}} \leftarrow \mathcal{V}_{\mathsf{uncover}} \setminus \mathsf{Candr}$\;
}
\Return $\mathcal{R}$\;
\end{small}
\end{algorithm}
\begin{algorithm}[t]
\label{AEUP}
\begin{small}
\caption{\textsc{FindUp}($v$)}
\LinesNumbered
\KwIn{the present violation feature $v$}
\KwOut{$\mathsf{Candr}$, \emph{UpSet}}
$\emph{Flag}[v] \leftarrow$ True\;
\If{CoverAE($v$) $\neq \emptyset $ }{
\Return $\emptyset$, $\emph{CoverAE}(v)$\;
}
Initialize $\mathsf{Candr} \leftarrow \{v \}, \emph{UpSet} \leftarrow \emptyset$\;
\ForEach{$v^* \in v.{\mathsf{neighbours}}$}{
\If{Flag$[v^*] =$ False }{
temp\_$\mathsf{Candr}$, temp\_\emph{UpSet} $ \leftarrow \textsc{FindUp}(v^*)$\;
$\mathsf{Candr} \leftarrow \mathsf{Candr} \cup \textrm{temp}\_\mathsf{Candr}$ \;
$\emph{UpSet} \leftarrow \emph{UpSet} \cup \textrm{temp}\_\emph{UpSet}$\;
}
}
\Return $\mathsf{Candr}$,\emph{UpSet}\;
\end{small}
\end{algorithm}
 \emph{Complexity}. In Algorithm \ref{update} lines 1-3, it costs $O(|\mathcal{V}|^2)$ to construct the graph, and  $O(|\mathcal{V}|\cdot |\mathcal{R}|)$ to find out the explanation set \emph{CoverAE} of all the features, where $|\mathcal{V}|$ and $|\mathcal{R}|$ denote the number of violation features and anomaly explanations, respectively. The outer loop (Lines 5-13) costs $O(\mathcal{V}_{\mathsf{uncover}})$ times, while the inner loop (Lines 9-12) spends $O(|\mathcal{V}_{\mathsf{cover}}|\cdot|\emph{UpSet}| + |\mathsf{Candr}| \cdot |\emph{UpSet}|)$. In practice, the size of $\mathcal{V}_{\mathsf{uncover}}$ becomes smaller with the outer loop. Thus, the outer loop is executed in $O(\frac{|\mathcal{V}_{\mathsf{uncover}}|}{|\mathsf{Candr}|})$ on average. Together, the whole loop costs $O\big(\frac{|\mathcal{V}_{\mathsf{uncover}}|}{|\mathsf{Candr}|} \cdot |\emph{UpSet}| \cdot \max\big \{ |\mathcal{V}_{\mathsf{cover}}|,|\mathsf{Candr}|\big\}\big)$. Whether $|\mathcal{V}_{\mathsf{uncover}}|$ is larger than $|\mathcal{V}_{\mathsf{cover}}|$ or not, it always has $|\emph{UpSet}|\leqslant |\mathcal{R}|$.
 To put it together, Algorithm \ref{update} totally costs $O\big(|\mathcal{V}|\cdot \max\{|\mathcal{V}|, |\mathcal{R}|\}\big)$ to update the whole knowledge set.
\newline\indent We highlight that, $r^+$ is considered to be inserted into the possible explanation set of a fault with a probability $w$, for the reason that the updated anomaly representations are derived from multiple real detections, whose dependability is less than the existing knowledge. We will discuss our blueprint of how to modify the knowledge set in the iteration of detections and updates in Section \ref{sec5.2}, including the consideration of the weight $w$ and the violation features of $r$.
\subsection{Modification on the Knowledge Set}
\label{sec5.2}
Note again that $R(E)$  is divided into $R^*(E)$ and  $R^+(E)$, where we highly trust the representations in $R^*(E)$ while we consider the representations in $R^+(E)$ with uncertainty. Actually, the possible explanation set $R^+(E)$ likely comes from the learning result of real detections, \emph{ i.e.}, the update process in Algorithm \ref{update}.
The violation features, especially the ones appearing frequently in detection phases, can be applied in the knowledge modification process.
\newline \indent  It is practicable to return anomaly explanation results to domain experts who are able to adjust and improve the existing knowledge system. And the improved knowledge would in turn contributes to the accuracy and the reliability of the following detections. Besides the  supplementary possible representations $r^+$s, we can also make the representation $r$ more accurate by modifying the degree function value $F(c)$ of $r$. In this section, we further discuss two kinds of potential knowledge modification strategies.
\newline \indent  \underline{\emph{Type-1 Modification: degree function values}}.
As proposed in Definition \ref{Vfeature}, the function value $F(c)$ of a violation feature $v$ measures to which degree a sequence violates the constraint $c$. For a violation feature of quantitative constraints denoted by $v: \langle S, F(c)=[d,u]\rangle$, let $F'(c)= [d',u']$ be the present degree function obtained from $k$ times of {updates}. We modify the function $F(c)$ \emph{w.r.t} $v$ as follows.
\begin{equation}\label{boolean}
  F(c) \left \{
\begin{array}{cl}
  = [d'',u''] = [\frac{k\cdot d'+d}{k+1}, \frac{k\cdot u' +u}{k+1}], & dist(F,F') < 1
  \\
  \textrm{will be returned to manual}, & dist(F,F') = 1.\nonumber \\
\end{array}
\right.
\end{equation}
\newline \indent \underline{\emph{Type-2 Modification: weights}}. The weight $w$ of a possible representation $r^+$ in an anomaly explanation $R$  would be estimated by the conditional probability $\Pr(r^+|R)$ in Equation (\ref{weights}),
\begin{equation}\label{weights}
  \hat{w}(r^+)= \Pr(r^+|R) = \frac{\Pr(R,r^+)}{\Pr(R)} = \frac{N_{\mathsf{positive}(Rr^+)}}{N_{\mathsf{positive}(R)}}
\end{equation}
where $N_{\mathsf{positive}(R)}$ denotes the occurrence number of anomaly explanation $R$ in solution $H$ during times of learning process, while $N_{\mathsf{positive}(Rr^+)}$ denotes the occurrence number of the conditions that $R$ exists in $H$ and $\exists \, v$ is consistent with $r^+$.
\section{Experimental Study}
\label{sec6}
We now evaluate the experimental study of the proposed methods. All experiments run on a computer with 3.40 GHz Core i7 CPU and 32GB RAM.
\subsection{Experimental Settings}
\label{sec6.1}
\textbf{Data source}. We conduct our experiments on real-life industrial equipment data, named \textsl{FPP}, which has 80 sensors recording the working conditions of a fan-machine group from a large-scale \underline{f}ossil-fuel \underline{p}ower \underline{p}lant. We have analyzed data on more than 1620\emph{K} historical time points for 5 consecutive months with log files and functional documents. We report our  experimental results on 64 sensors after preprocessing.
\newline\indent \textbf{Implementation}.  We have developed \emph{Cleanits}, a data \underline{clean}ing system for \underline{i}ndustrial \underline{t}ime \underline{s}eries in our previous work \cite{DBLP:journals/pvldb/DingWSLLG19}, which reads and writes data from Apache IoTDB \cite{DBLP:journals/pvldb/0018HQ00MFZ0ZKJ20}. The anomaly explanation method proposed in this paper is applied as one main function of \emph{Cleanits}.
\newline \indent We implement all algorithms proposed in this paper, with the constraint-based detection method $\mathsf{VioDetect}$, the explanation algorithm $\mathsf{AEC}$, and the update algorithm $\mathsf{Update}$. The constraints used in experiments contain half \emph{real} constraints provided by domain knowledge and half \emph{synthetic} ones concluded by a long-term research of both the historical data and log files. Due to the fact that not only the industrial knowledge set is far from comprehensive, but also the labelled anomalies as well as explanations are limited, we extend the
existing knowledge (mostly from documents and domain experts), and manually regulate some synthetic explanations with the corresponding representations based on the acknowledge documents and fault logs.
\newline \indent We consider the original \emph{clean} time series data as ground truth, and inject anomalies \emph{w.r.t} constraints into sequences in different time intervals. Without loss of generality, we introduce anomaly instances according to the error types in \cite{2010Outliers, 1998Outliers}, and deeply consider the anomaly patterns simultaneously located in multiple sequences referring to the acknowledged real anomaly events. We totally apply 210 constraints and 60 anomaly events as the given knowledge set.
\newline \indent Besides the porposed method, we also implement five algorithms for comparative evaluation:
\begin{itemize}
\item $\mathsf{greedyC}$: uses greedy strategy for the set cover problem in the candidate explanation set $G$ to iteratively select event $E$ satisfying $R(E) = \arg\min \frac{Cost(E)}{|R.cover \cap \mathcal{V}_{\mathsf{uncover}}^*|}$, with constraint types insensitive.
\item $\mathsf{greedynC}$: describes the violation in multi-sequence constraints $c$ with $n$ violation features in the involved $n$ sequences \emph{w.r.t} $c$, rather than apply only one feature to denote such violations, with others the same with $\mathsf{greedyC}$.
    \item $\mathsf{MFnC}$: treats the multi-sequence-constraint violation with $n$ features in the involved sequences \emph{w.r.t} $c$, with others the same with the proposed $\mathsf{AEC}$.
    \item $\mathsf{TopK}$: sorts the explanations $R(E)$s in the ascending order of $Cost(E)$, and chooses the top $K$ explanations as the result. $K$ is determined by $K=|\mathcal{R}|\cdot \frac{|\mathcal{C}_{\textsf{vio}}|}{|\mathcal{C}|}$, where $|\mathcal{C}|$ is the number of adapted constraints and $|\mathcal{C}_{\textsf{vio}}|$ is the number of detected violated constraints.
        \item $\mathsf{AE}$: output all explanations satisfying $R(E) = \arg \frac{Cost(E)}{|R(E).cover|} \leq\lambda$, where $\lambda\in(0,1)$ is a set threshold. We report results with $\lambda$=0.4, for it provides the best results among possible threshold values.
\end{itemize}
We note that the first three algorithms are cover-based, while  $\mathsf{TopK}$ and $\mathsf{AE}$ are not.
\newline \indent \textbf{Measure}.
We apply Precision ($\mathsf{P}$) and  Recall ($\mathsf{R}$) metrics to evaluate the performance of algorithms in Equation (\ref{PR}). $\mathsf{P}$ measures the ratio between the number of correctly-identified anomaly events, \emph{i.e.,}  $\textrm{\#}{\textsl{correctIdentifyAE}}$, and the total number of anomaly events identified by algorithms. $\mathsf{R}$ is the ratio between $\textrm{\#}{\textsl{correctIdentifyAE}}$ and the total number of anomaly events which actually happened.
\begin{equation}\label{PR}
\mathsf{P}= \frac{\textrm{\#}{\textsl{correctIdentifyAE}}}{\textrm{\#}\textsl{IdentifyAE}}, \mathsf{R}= \frac{\textrm{\#}\textsl{correctIdentifyAE}}{\textrm{\#}\textsl{AE}}.
\end{equation}
\subsection{General Performance}
\label{6.2}
With the condition that \#\emph{Constraints}=210, we perform all comparison algorithms on 4 datasets {which has 10.8K time points on 64 sensors recording data for one day}. About 20 anomaly events occur in each dataset. As shown in Figure \ref{e1}, $\mathsf{VioDetect}$ reaches high performance on both $\mathsf{P}$ and $\mathsf{R}$ on all datasets. This is the foundation for a high-quality explanation computing. The proposed $\mathsf{AEC}$ has the best scores with
$\mathsf{P}=0.87$ on average, when $\mathsf{MFnC}$ comes the second. It reveals that it is better to treat the violation of a multi-sequence constraint as only one violation feature, than maintain $n$ features in each involved sequence. The gap in $\mathsf{P}$ between $\mathsf{greedyC}$ and $\mathsf{greedynC}$ also confirms this. However, both
algorithms fail to provide precise explanations. It is because they both treat all types of constraints equally, and fail to give priority to multi-sequence constraints, whose violations are more likely to show major features to be identified as an anomaly event.
\newline \indent Figure \ref{e1} shows the four cover-based algorithms have similar Recall scores in the four datasets. It reveals that the covering solutions can capture and recall at least 85\% of the anomaly events. In addition, it is almost no difference in Recall whether the constraint types are sensitive or not. As for algorithm $\mathsf{TopK}$ and $\mathsf{AE}$, the performance of both algorithms is not steady in different datasets. This shows that simply choosing explanations \emph{w.r.t} $Cost$ cannot well identify the occurred anomaly events.
\begin{figure}[t]
\centering
 {
\includegraphics[width=1.673in]{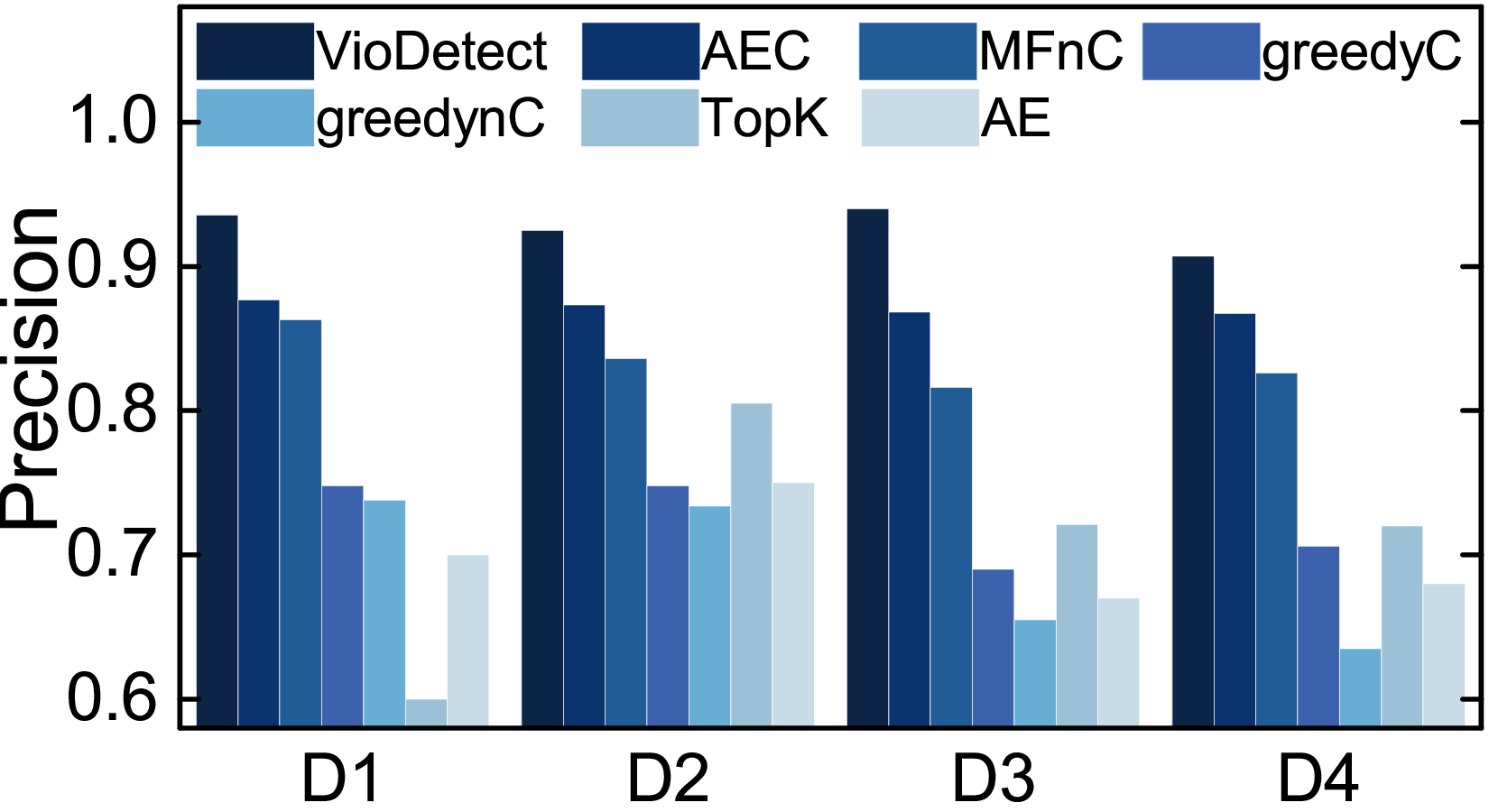}}
{
\includegraphics[width=1.673in]{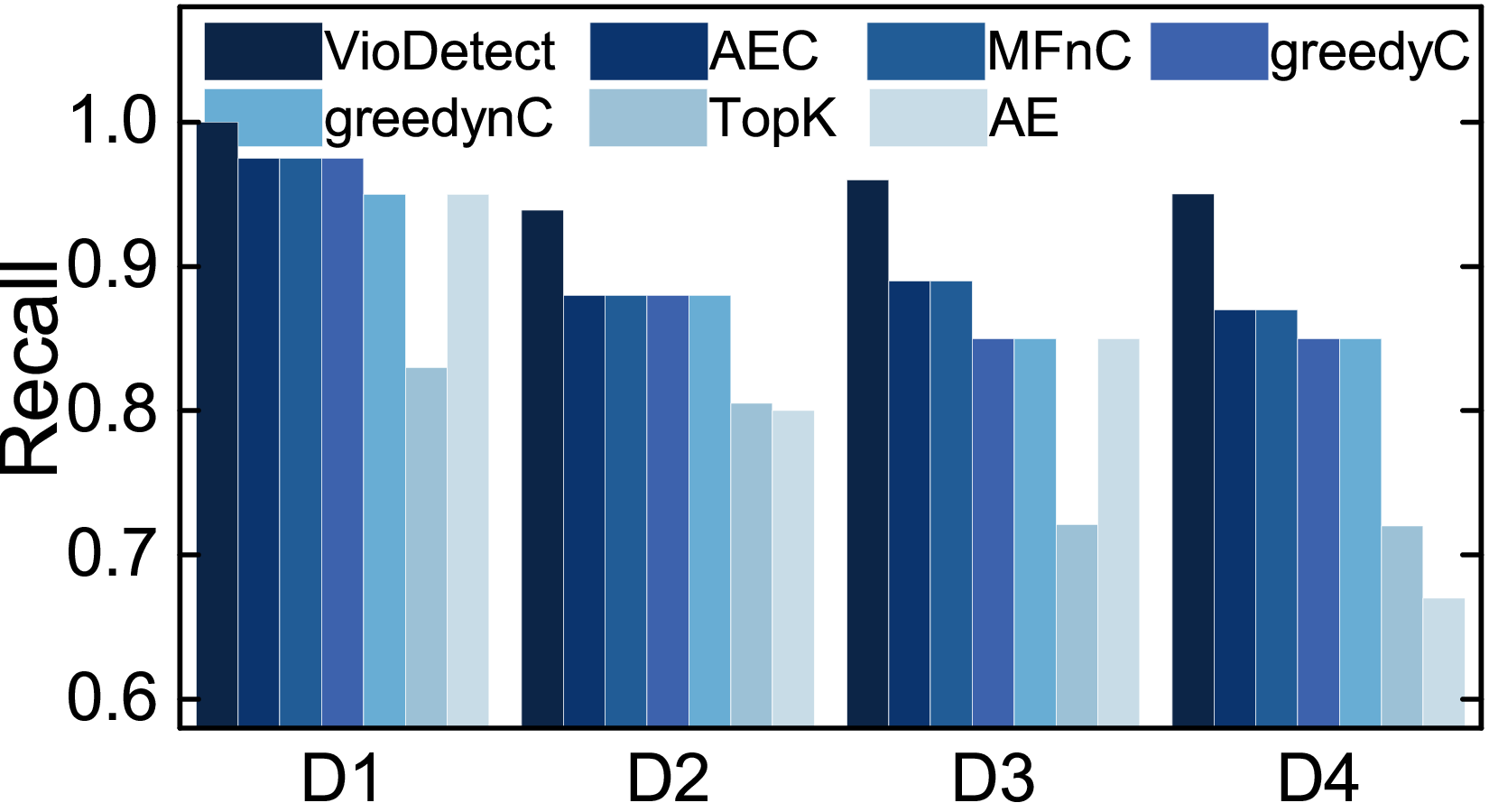}}
\vspace{-0.8em}
\caption{General performance comparison on 4 datasets in \textsl{FPP}.}
\label{e1}
\end{figure}
\subsection{Evaluation on Explanation Performance}
\label{6.3}
\begin{figure}[t]
\centering
 {
\includegraphics[width=1.45in]{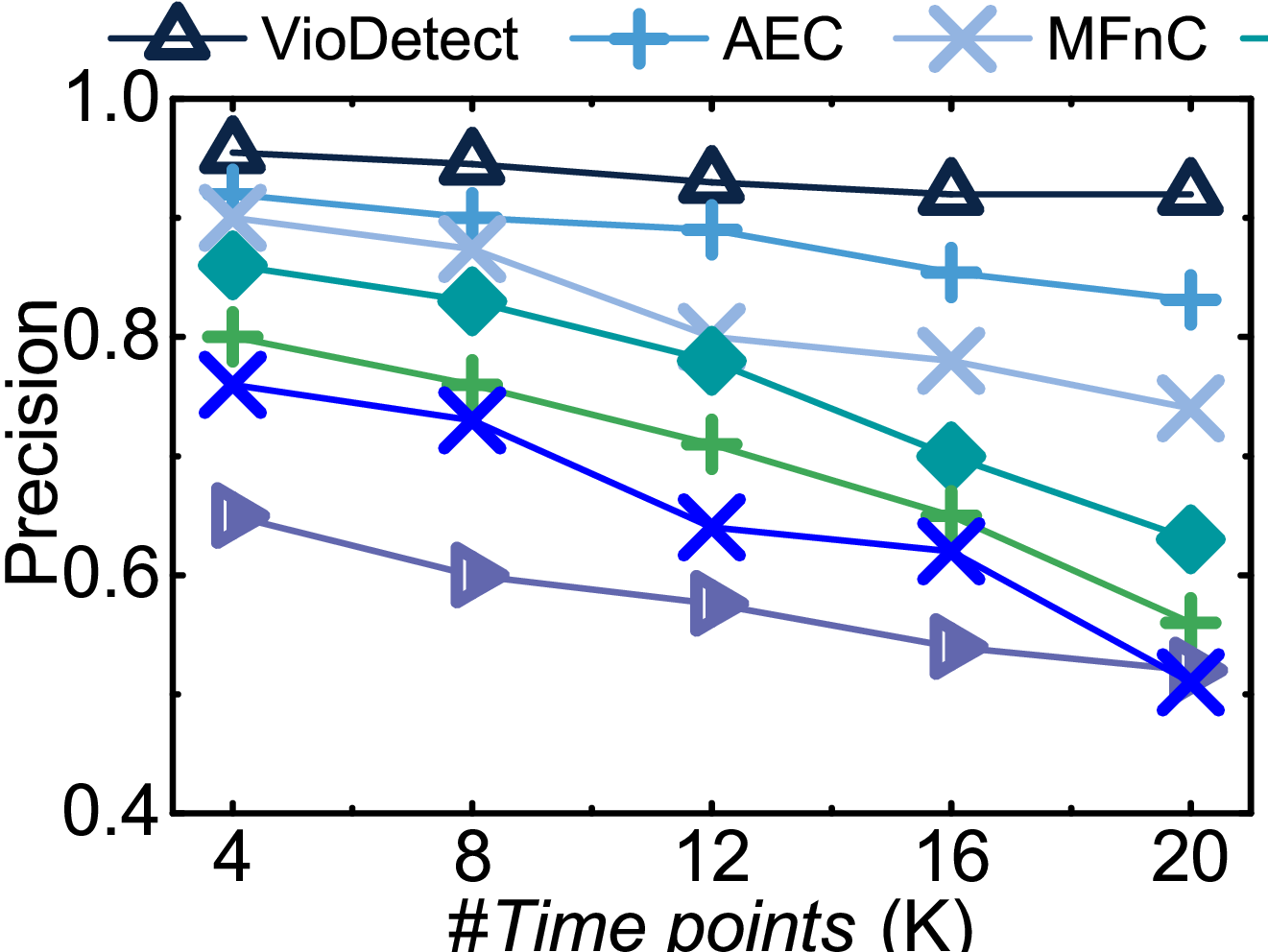}}
{
\includegraphics[width=1.45in]{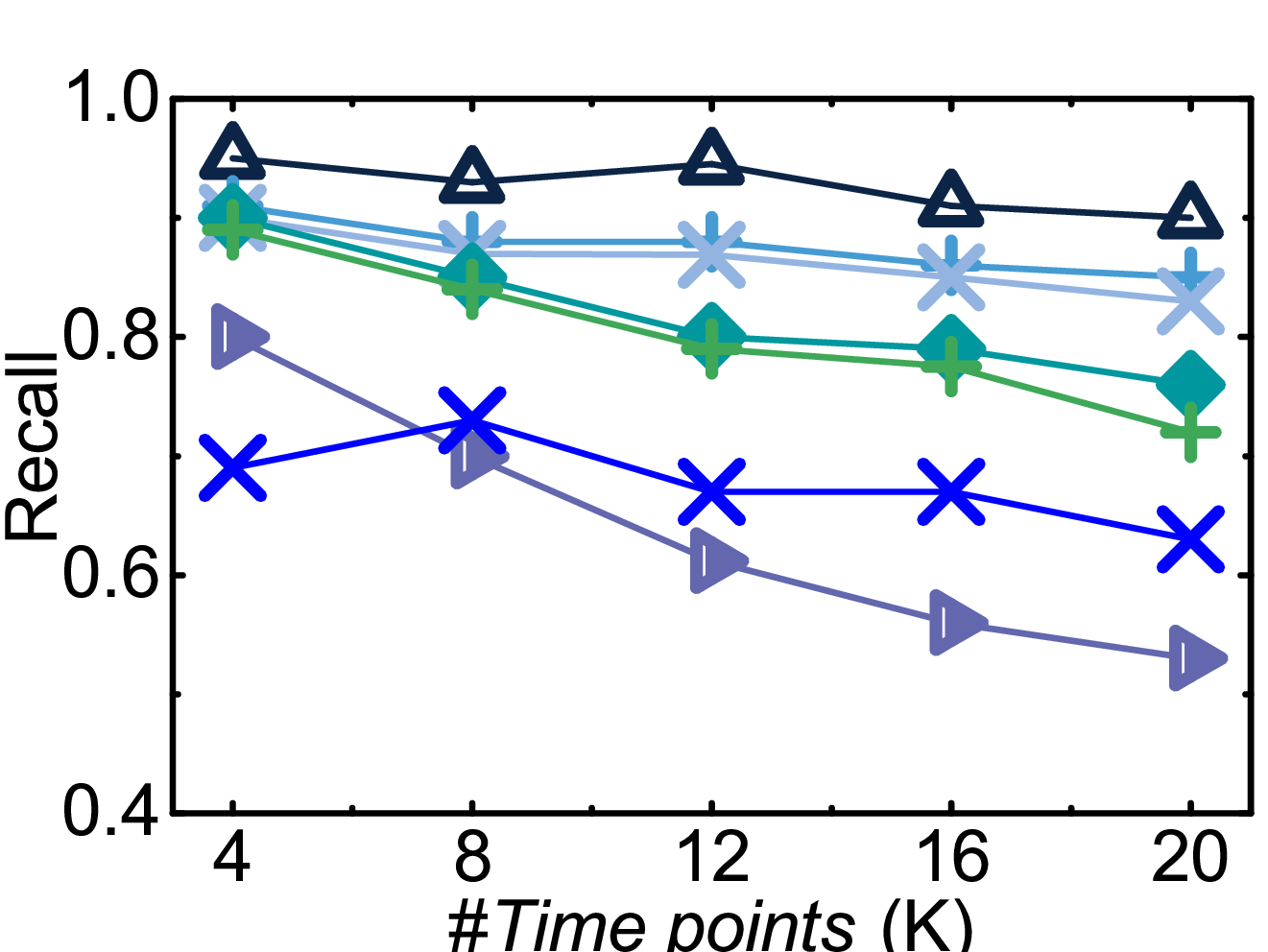}}
\vspace{-0.8em}
\caption{Performance comparison vs. the number of time points}
\label{e3}
\end{figure}

\begin{figure}[t]
\centering
 {
\includegraphics[width=1.45in]{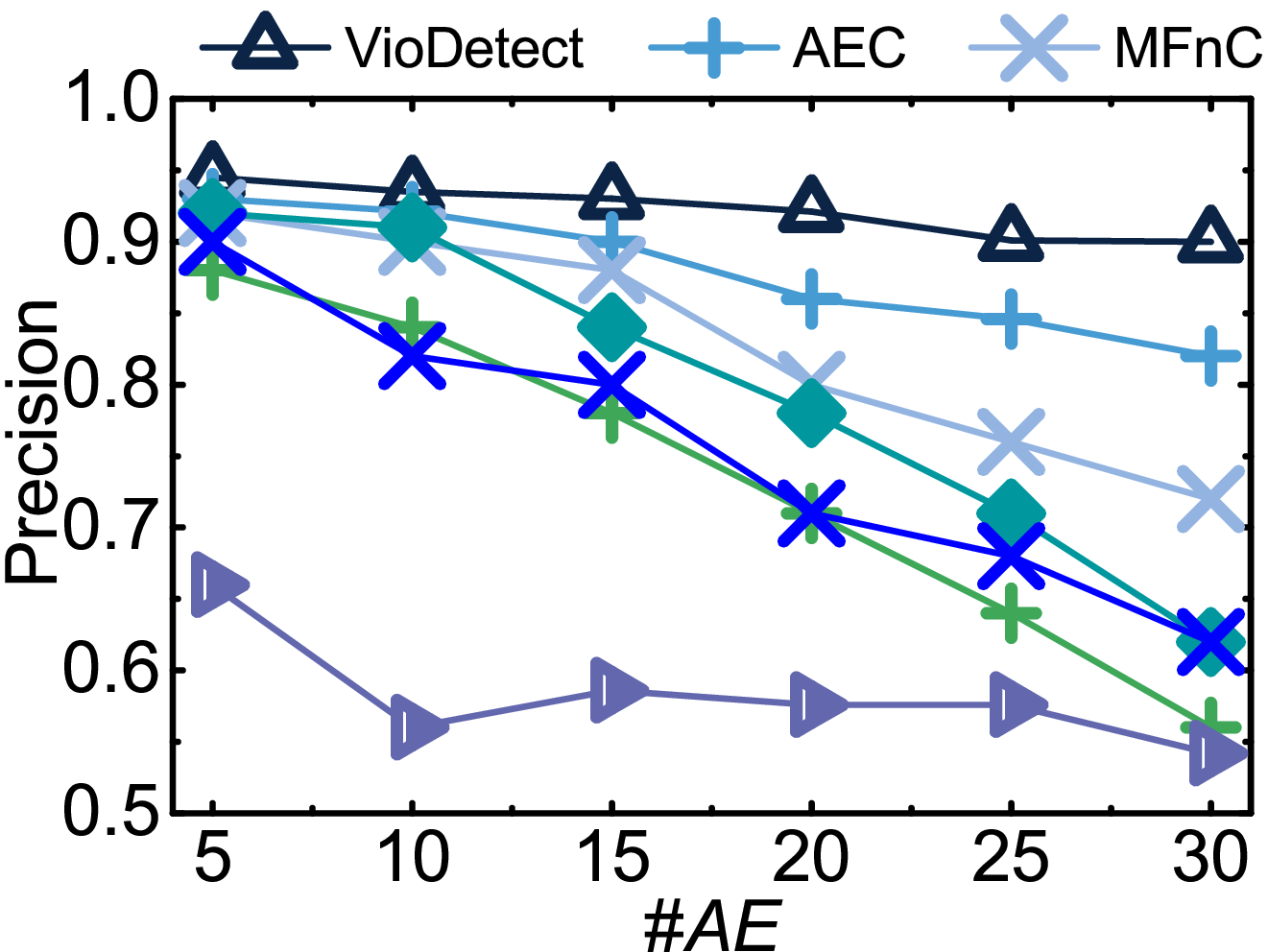}}
{
\includegraphics[width=1.45in]{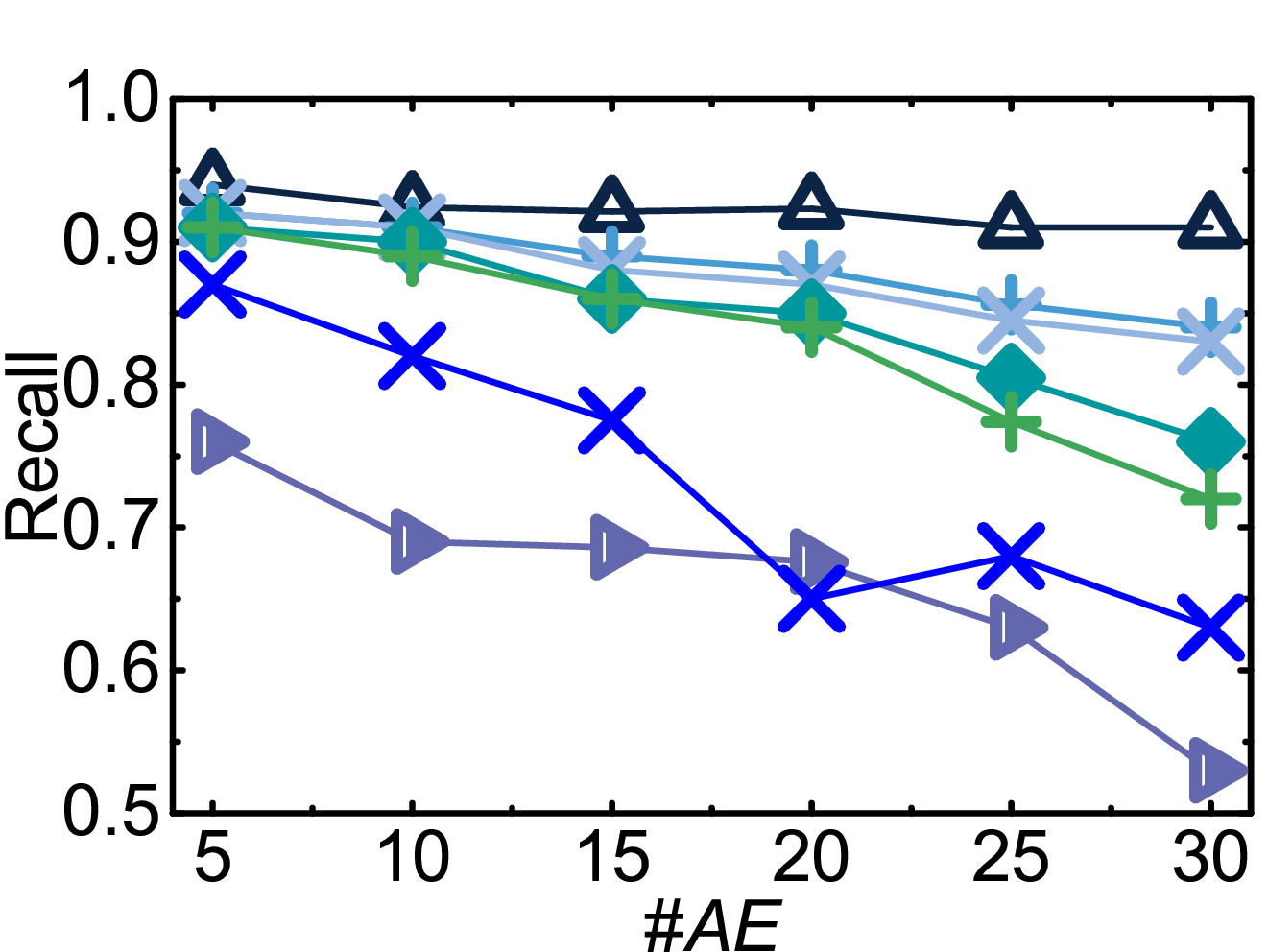}}
\vspace{-0.8em}
\caption{Performance comparison vs. the number of anomaly events}
\label{e4}
\vspace{-1.2em}    
\end{figure}

We next report the performance under three vital parameters: constraints amount (\#\emph{Constraints}), data amount (\#\emph{Time points}), and the number of occurred anomaly events (\#\emph{AE}).
\newline \indent \textbf{Varying \#\emph{Constraints}}.
Figure \ref{e2} shows the performance on the condition that \#\emph{Time points}=20K. It shows that $\mathsf{VioDetect}$ and the four cover-based algorithms have higher scores with the increasing of \#\emph{Constraints}, which indicates the completeness of the constraint set affects the explaining quality.
\newline \indent Both $\mathsf{P}$ and $\mathsf{R}$ of the cover-based algorithms trend to be stable after \#\emph{Constraints} reaches 120. Our proposed $\mathsf{AEC}$ shows the best scores on $\mathsf{P}$ ($>$0.85) and $\mathsf{R}$ ($>$0.86), with $\mathsf{MFnC}$ comes the second, and $\mathsf{greedyC}$ the third. It also shows that the precision difference between $\mathsf{AEC}$ and $\mathsf{MFnC}$ becomes larger with the increasing constraint amount. It verifies again that the advantage of describing the violation \emph{w.r.t} one constraint with only one feature. As for Recall, $\mathsf{AEC}$ and $\mathsf{MFnC}$ (resp. $\mathsf{greedyC}$ and $\mathsf{greedynC}$) present quite closed scores, which shows that the constraint-type sensitive does not obviously effect the Recall level.
\begin{figure}[t]
\centering
 {
\includegraphics[width=1.45in]{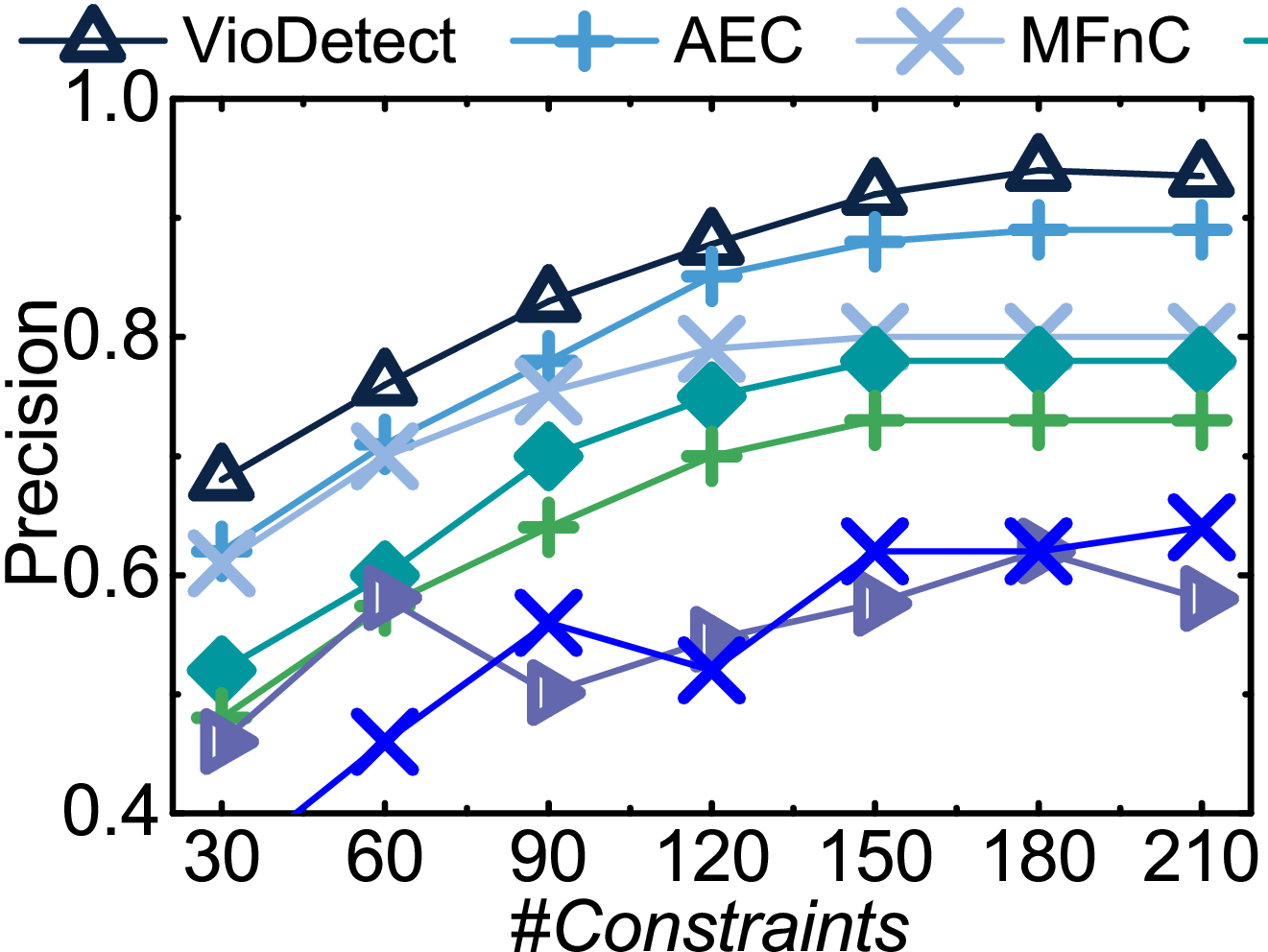}}
{
\includegraphics[width=1.45in]{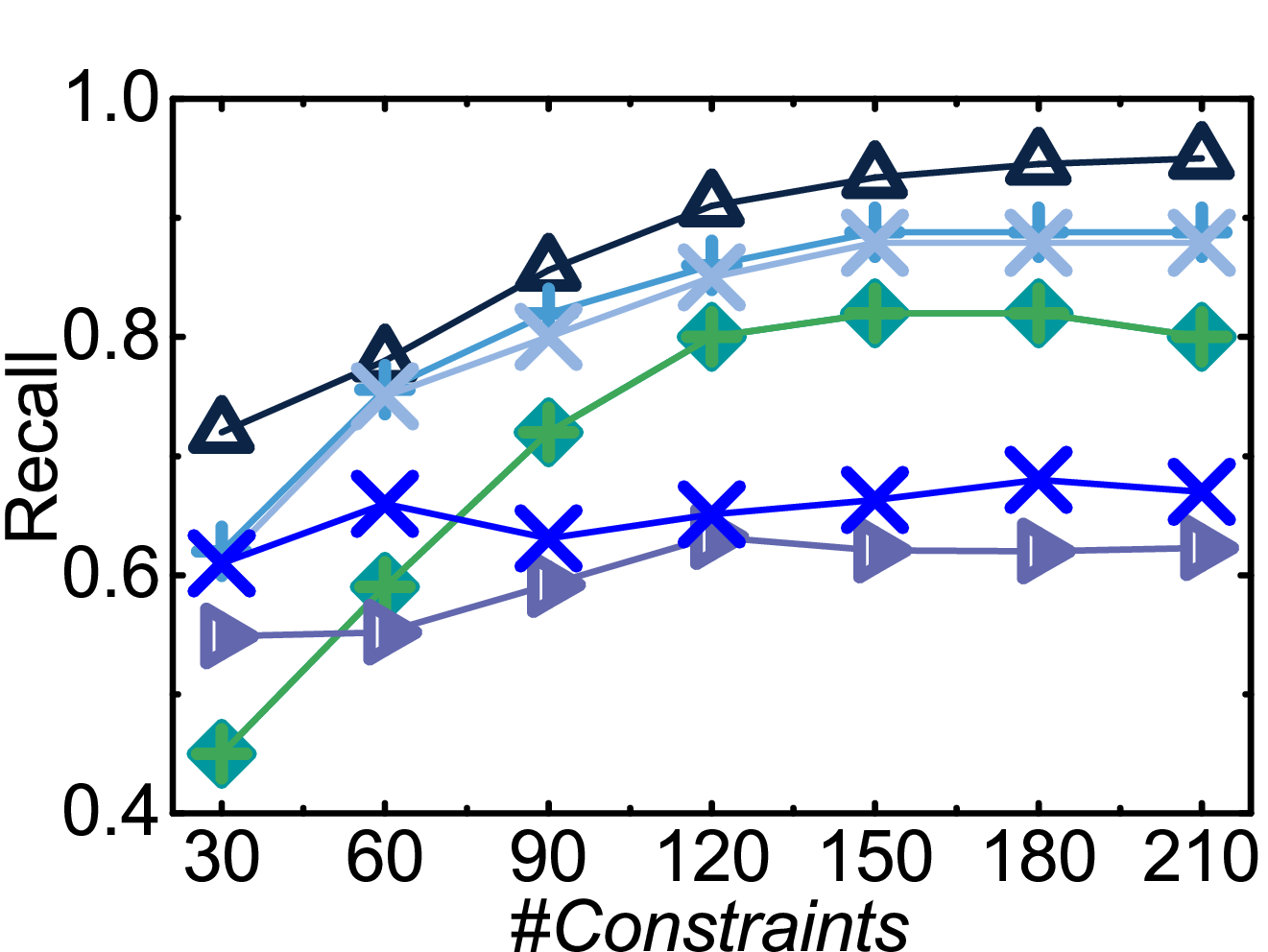}}
\vspace{-0.8em}
\caption{Performance comparison vs. the number of constraints}
\label{e2}
\vspace{-1.2em}    
\end{figure}
\newline\indent \textbf{Varying \#\emph{Time points}}.
Figure \ref{e3} shows results with \#\emph{Constraints}=210 and varying \#\emph{Time points}. Anomaly events are evenly located in data. The detection performance of  $\mathsf{VioDetect}$ is stable and only has a little drop when the data amount becomes larger, and it achieves $\mathsf{P}$=0.92 and $\mathsf{R}$=0.91 in detecting almost 2-day data (\emph{i.e.}, \#\emph{Time points}=20K).
\newline \indent While all the comparison algorithms have drops in different degree with the increasing data amount, $\mathsf{AEC}$ beats the rest comparison algorithms in both metrics. It achieves $\mathsf{P}>$0.83 and $\mathsf{R}>$0.85 with \#\emph{Time points}$\leq$20K. It is found that the falling speed of $\mathsf{greedyC}$ and $\mathsf{greedynC}$ is faster. It shows that the simple greedy strategy cannot provide good solutions when data amount gets larger, where exists more anomaly instances.
\newline \indent \textbf{Varying \#\emph{AE}}.
Figure \ref{e4} reports the results with \#\emph{Constraints}=210, and varying number of happened anomaly events in \#\emph{Time points}=20K data. $\mathsf{VioDetect}$ has stable Recall scores while Precision has a little drop from 0.942 to 0.904 with \#\emph{AE} varying from 5 to 30. It shows that the anomaly event amount does not obviously effect the the violation detection result.
\newline \indent $\mathsf{AEC}$ has the least drop in both metrics among the five comparison methods, and presents $\mathsf{P}>$0.825 and $\mathsf{R}>$0.846 when \#\emph{AE} reaches 30. It confirms the effectiveness of our method when faced with quite a few anomaly events. The fast decline in the performance of $\mathsf{greedyC}$ and $\mathsf{greedynC}$ shows again that the naive greedy-based algorithms would compute less reliable solutions when the growth number of anomalies. The baseline algorithm $\mathsf{TopK}$ also drops with the increasing \#\emph{AE}, and it always provides poor results compared with others. Note that another baseline algorithm $\mathsf{AE}$ has closer gaps with the cover-based algorithms in $\mathsf{P}$, however it has poor Recalls.
\newline \indent We highlight that \#\emph{AE}$>$20 is a strict condition in real scenarios, even through in a 2-day time series data. Thus, the higher performance of our proposed $\mathsf{AEC}$ as well as the large gap between $\mathsf{AEC}$ and others algorithms indicates
our method has the potential of the effectiveness and robustness in solving real anomaly explanation problems.
\subsection{Evaluation on Update Performance}
\label{6.4}
\begin{figure}[t]
\centering
 {
\includegraphics[width=1.3in]{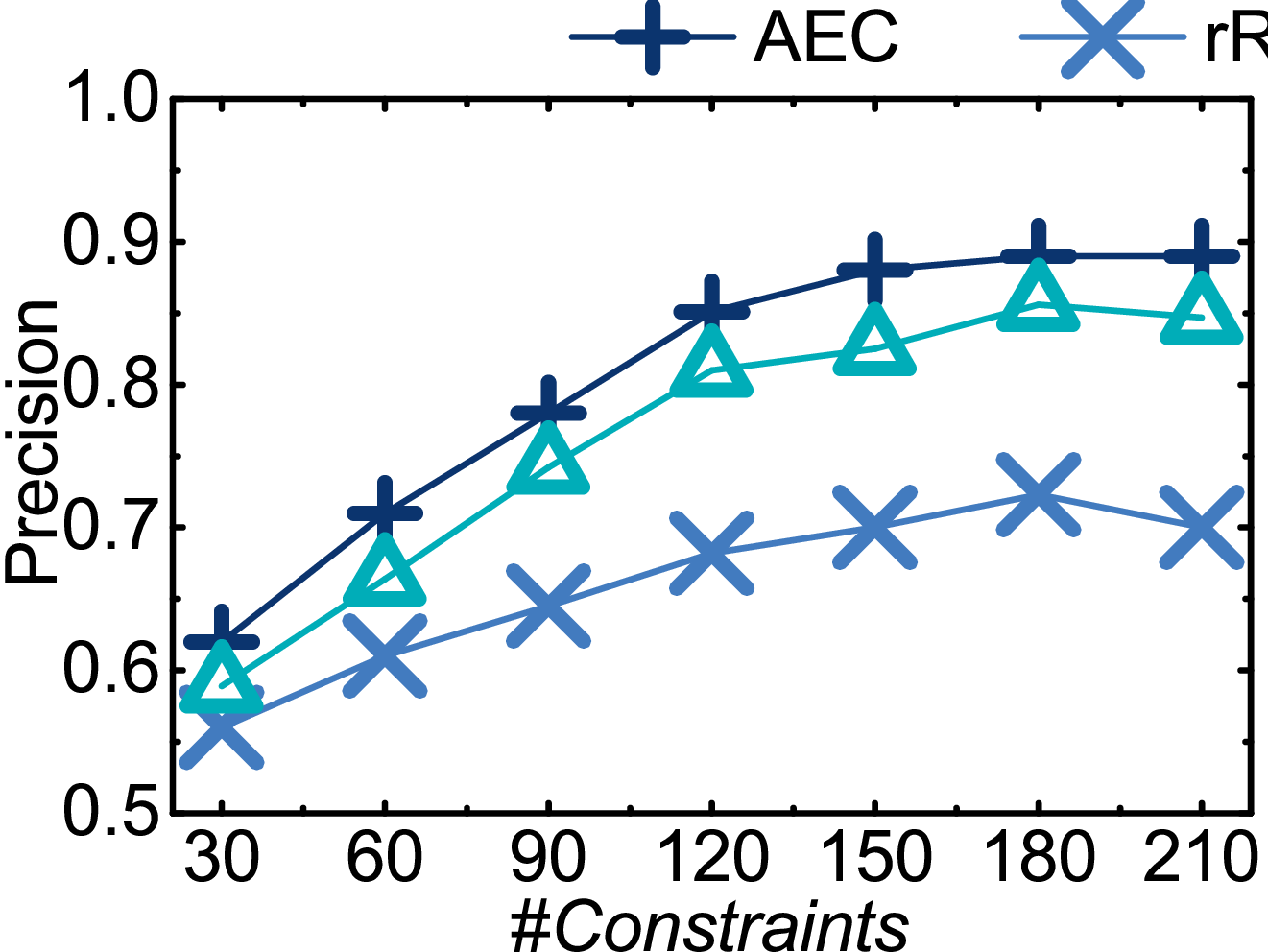}}
{
\includegraphics[width=1.3in]{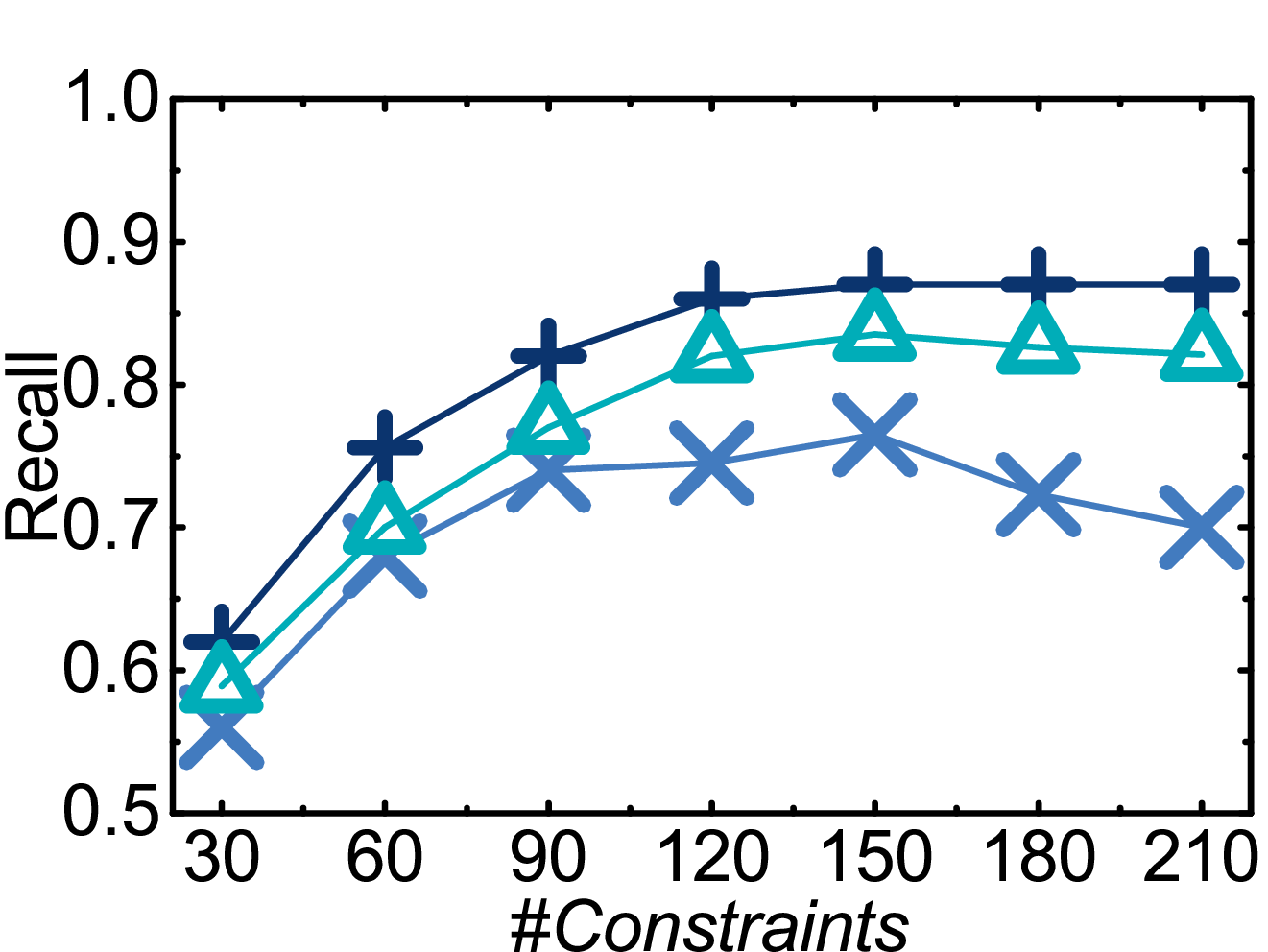}}
\vspace{-0.8em}
\caption{Update performance vs. the number of constraints}
\label{e5}
\vspace{-1.2em}    
\end{figure}

\begin{figure}[t]
\centering
 {
\includegraphics[width=1.3in]{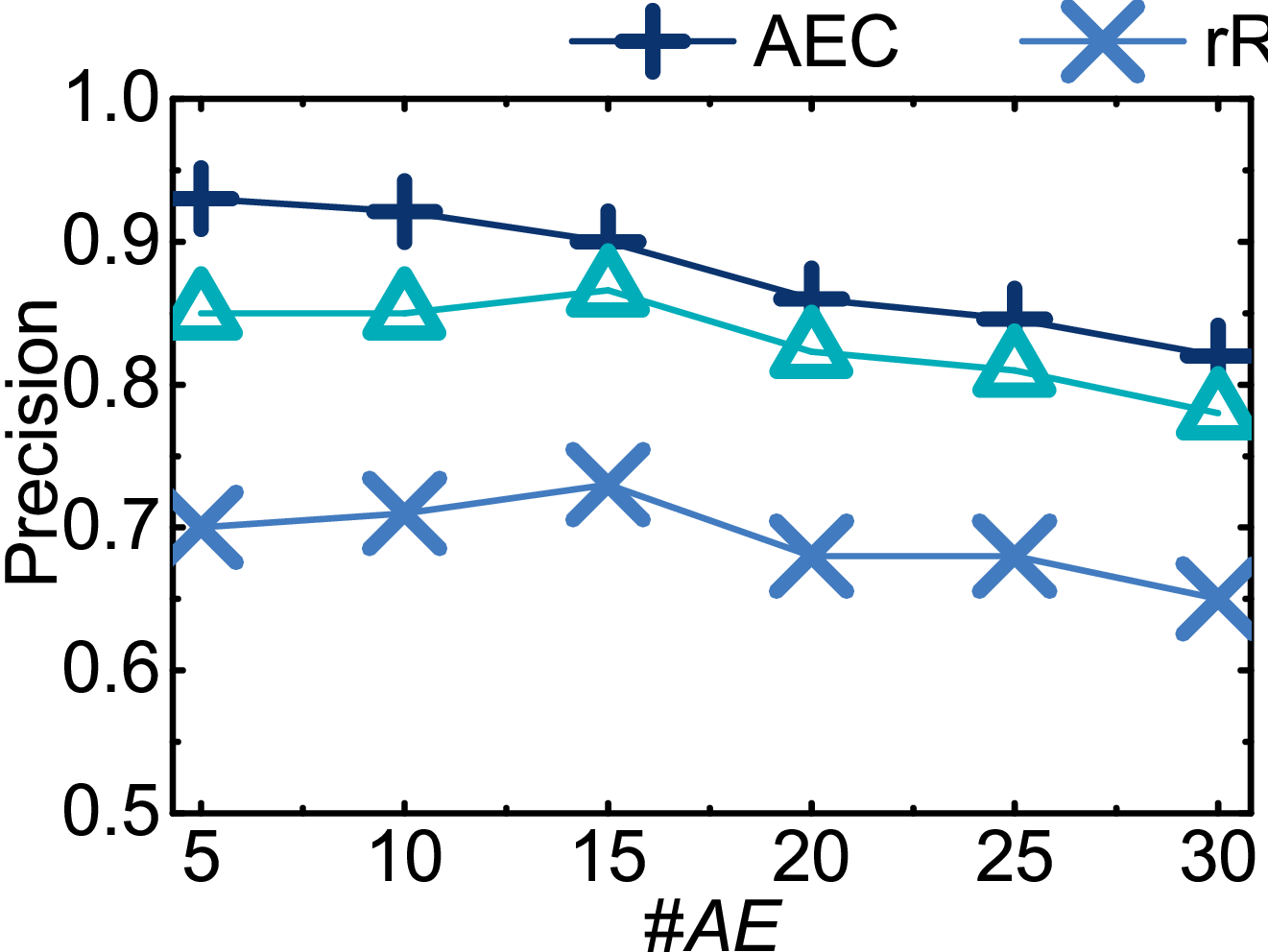}}
{
\includegraphics[width=1.3in]{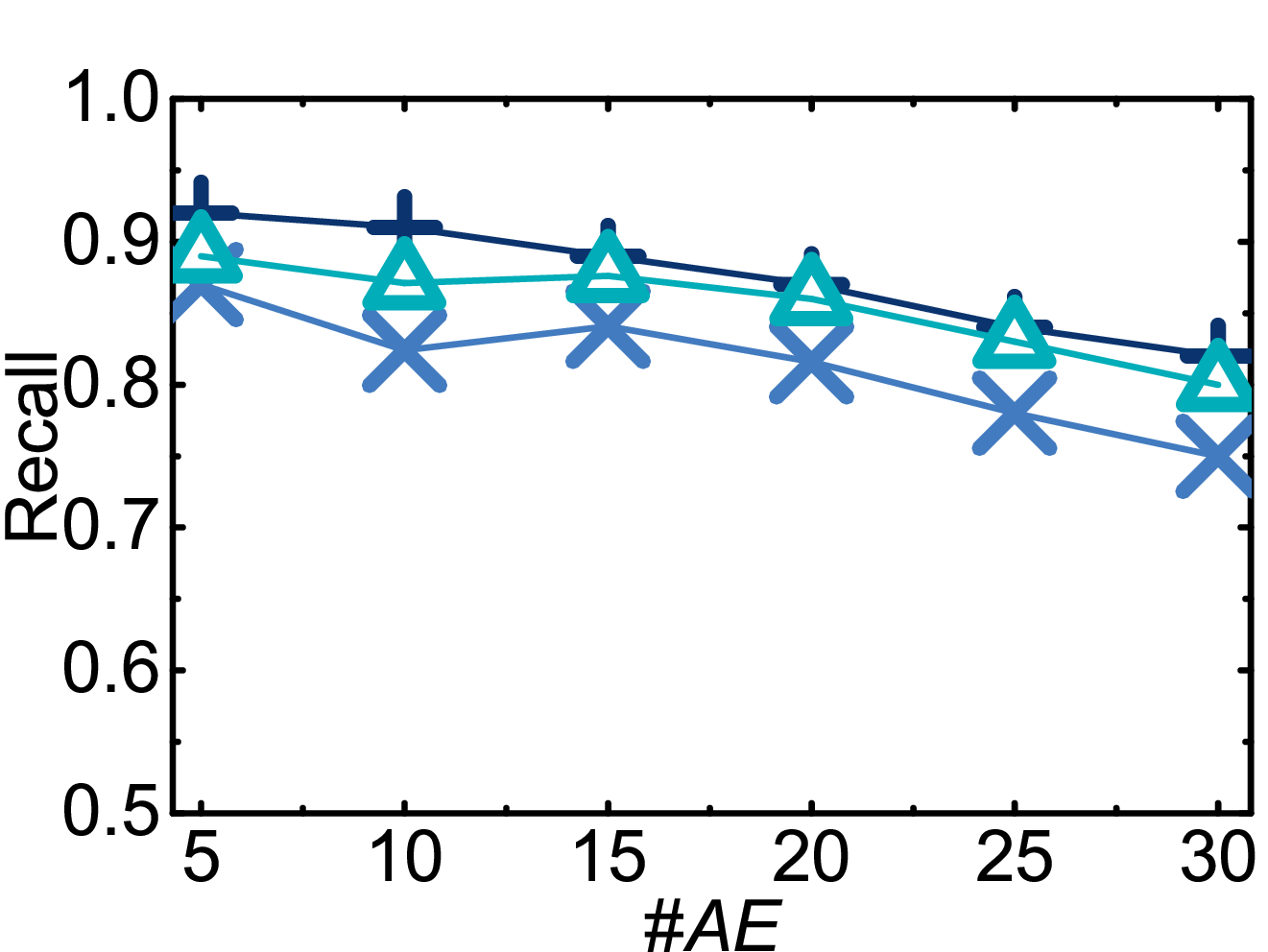}}
\vspace{-0.8em}
\caption{Update performance vs. the number of anomaly events}
\label{e6}
\vspace{-1.5em}    
\end{figure}

We then introduce the method performance of our knowledge update phase with three parameters: \#\emph{Constraints}, \#\emph{AE}, and the incomplete rate in explanations of the knowledge set $\mathcal{R}$ (\emph{inr \%}). We treat the performance of $\mathsf{AEC}$ with
the original $\mathcal{R}$ as baseline. We randomly select the explanation set of some anomaly events in $\mathcal{R}$, where we delete a percentage of possible representations. We denote the knowledge set after deletion as $\mathcal{R}^-$, and report the explanation results with $\mathcal{R}^-$ by $\mathsf{rRemove}$. We execute the update algorithms proposed in Section \ref{sec5} for $\mathcal{R}^-$, and denote the updated knowledge set as $\mathcal{R}^+$. We report the explanation results with $\mathcal{R}^+$ by $\mathsf{Update}$.
\newline \indent \textbf{Varying \#\emph{Constraints}}.
Figure \ref{e5} shows the performance under $\mathcal{R}$, $\mathcal{R}^-$, and $\mathcal{R}^+$ on the condition that \#\emph{Time points}=20K, and $\emph{inr} \%$=15\%. $\mathsf{rRemove}$ has the worst scores on both $\mathsf{P}$ and $R$, and the score gap between $\mathsf{AEC}$ and $\mathsf{rRemove}$ trends to be larger with the growth of  \#\emph{Constraints}. It shows that the incomplete knowledge set would reduce the quality of anomaly explanation results. When it comes to $\mathsf{Update}$ with $\mathcal{R}^+$, it shows a significant improvement on both metrics. Our update method assists to recover 93\% of the original performance with $\mathcal{R}$ on average. Moveover, the update result becomes better with the increasing constraint numbers. It is because the proposed Algorithm \ref{AEUP} effectively computes the relevance between uncovered violation features and the existing representations, which helps the recovery and update of the incomplete knowledge set, and further contributes to the improvement of explaining the anomalies.
\newline \indent \textbf{Varying \#\emph{AE}}.
Figure \ref{e6} presents the performance on the condition \#\emph{Constraints}=210, \#\emph{Time points}=20K, and \emph{inr}\%=15\%. The Precision scores of $\mathsf{rRemove}$ and $\mathsf{Update}$ are stable against the growth in \#\emph{AE}, while the Recall scores have an obvious drop. However, the performance differences of $\mathsf{rRemove}$ and $\mathsf{Update}$ in $\mathsf{P}$ is quite larger than the differences in $R$. While $\mathsf{rRemove}$ only reaches $\mathsf{P}$=0.7 on average, $\mathsf{Update}$ is able to provide more precise results and reaches 0.8 in $\mathsf{P}$ faced with 30 anomaly events. $\mathsf{Update}$ recovers 94\% of $\mathcal{R}$ on average. As for Recall, results on the three knowledge sets have similar scores in $\mathsf{R}$, this indicates, to some degree, our anomaly explanation method has robustness in Recall against an incomplete knowledge set.
\newline \indent \textbf{Varying Incomplete rate in $\mathcal{R}$}.
\begin{figure}[t]
\centering
{
\includegraphics[width=1.3in]{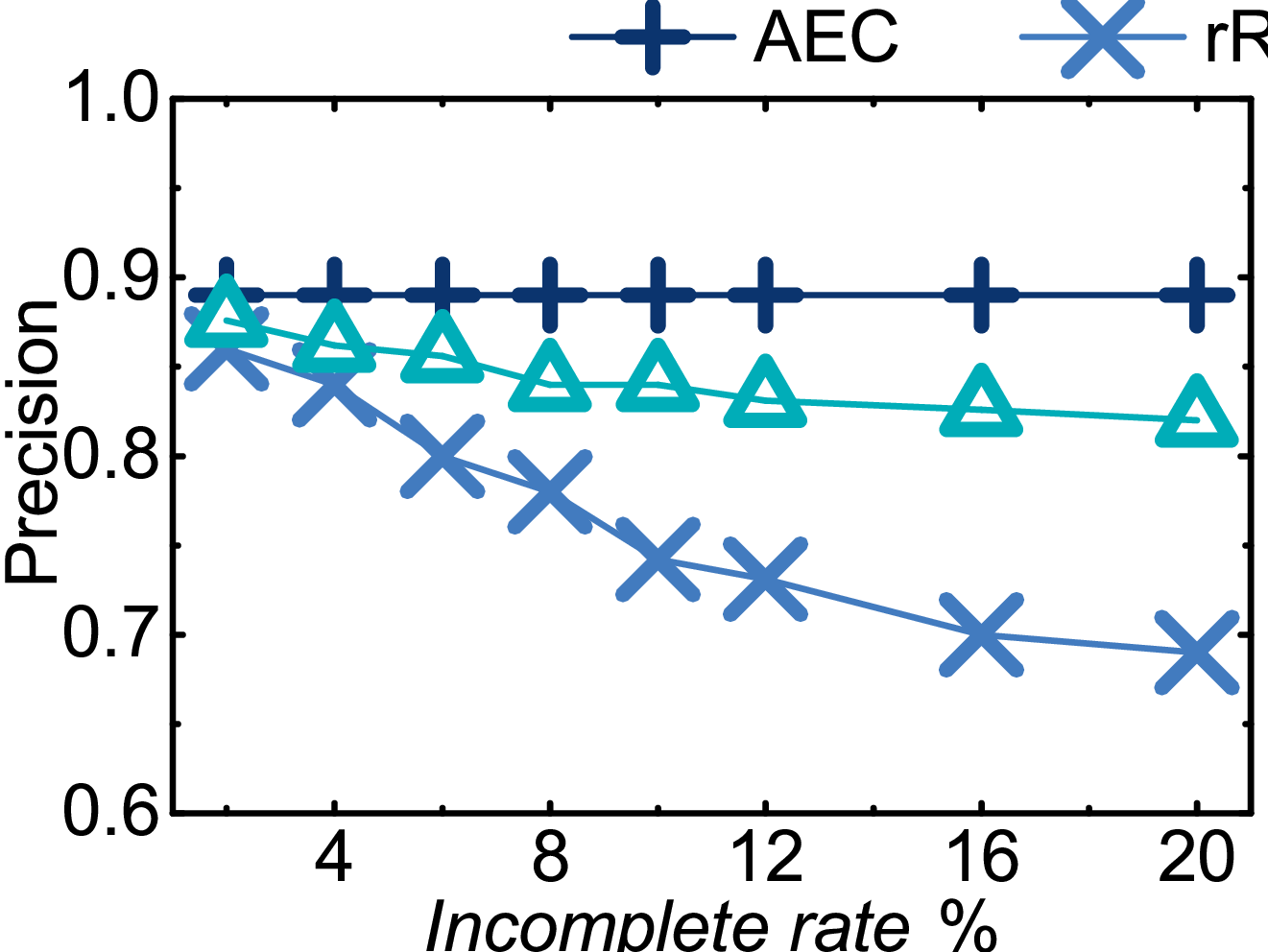}}
{
\includegraphics[width=1.3in]{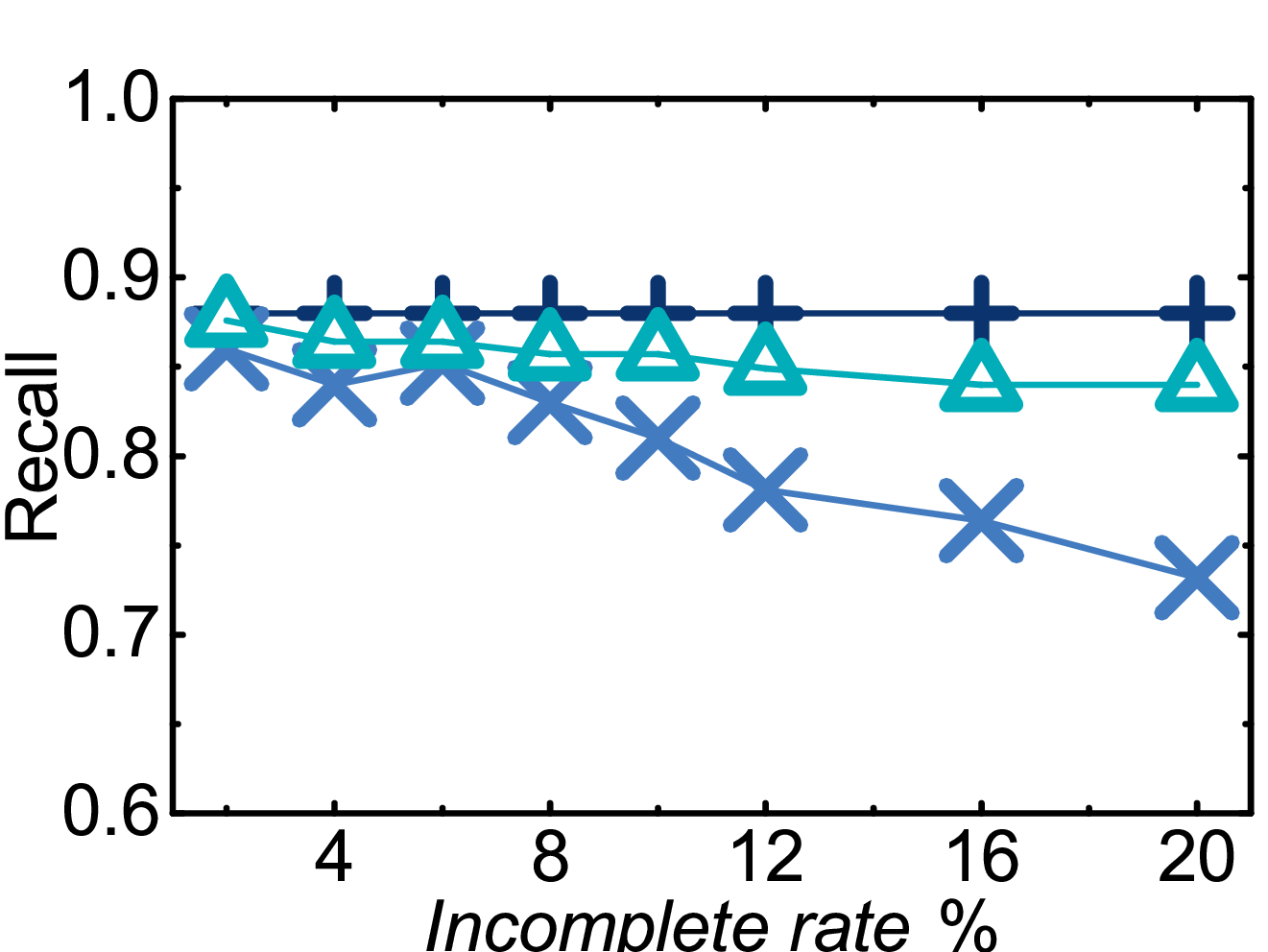}}
\vspace{-0.8em}
\caption{Update performance vs. the incomplete rate of $\mathcal{R}$}
\label{e7}
\vspace{-1.5em}    
\end{figure}
Figure \ref{e7} shows the results on the condition \#\emph{Time points}=20K and \#\emph{Constraints}=210 with varying incomplete rate of $\mathcal{R}$. We put $\mathsf{P}$=0.89 and $R$=0.88 of $\mathsf{AEC}$ with the original $\mathcal{R}$ along the X-axis, and focus on the method performance on $\mathcal{R}^-$ and $\mathcal{R}^+$. It is obvious that the scores of both metrics drop with the increasing \emph{inr}\%. It presents only 0.69 in $\mathsf{P}$ and 0.73 in $\mathsf{R}$ with the 20\%-incomplete knowledge set. Our update method is able to correctly recover the missing representations by computing and analyzing the uncovered violation features. $\mathsf{Update}$ recovers to 96.8 in $\mathsf{P}$ and 0.9881 in $\mathsf{R}$ of $\mathsf{AEC}$ with \emph{inr}\%=4\%, and 0.921 in $\mathsf{P}$ and 0.954 in $\mathsf{R}$ with \emph{inr}\%=20\%.
\newline \indent Though $\mathsf{Update}$ achieves effective improvement of explaining anomalies with incomplete knowledge set, it does not perfectly recover to the same level of $\mathsf{AEC}$ with the original set $\mathcal{R}$. It is believed that the relation among anomaly representations within or between anomaly events is quite complex. When some representations are missing in the given knowledge, it is challenged to be automatically well-identified and well-recovered. The improvement between the performance gap of $\mathsf{AEC}$ and $\mathsf{Update}$ and  the gap between $\mathcal{R}$ and $\mathcal{R}^+$ will be addressed in our future work.
\subsection{Efficiency results}
\label{sec5.5}
We report the time costs of the methods in Table \ref{time}, which presents results with several typical parameter values due to the limited space. Here, In\_Round denotes the iteration times of the explanation and update. It is worth noting that the first two steps in our method (\emph{i.e.,} violation detection and anomaly explanation) spend little time (denoted as AE\_Time), and finish computing the anomalies in 2 seconds with 210 constraints on 20K data. The update time cost increases with the growth of \#\emph{Constraints} and In\_Round, respectively. It shows that \#\emph{AE} has slight effect on the update time, for the reason that the limited \#\emph{AE} in real scenarios will not lead to large computation in either the covering and the update algorithms. Though the iteration times lead time growth, the proposed method can present almost the same performance with less In\_Round. Such efficiency performance shows that our method has the potential to be used for processing large-scale IoT data.
\begin{table}
\centering
  \caption{Time costs (\#\emph{Time points}=20K, \emph{inr}\%=15\%) }
  \label{time}
  \vspace{-0.8em}
\scalebox{0.8}{  \begin{tabular}{|c|c|c|c|c|c|c|}
    \hline
     $|\mathcal{C}|$ &\#\textbf{\emph{AE}}  & \textbf{AE\_Time } & \textbf{AE\_F1 } &\textbf{In\_Round} & \textbf{UP\_Time}& \textbf{UP\_F1 } \\
    \hline
    60 & 20 & $<$1s & 0.715& 3K & 4.69s & 0.656\\
     \hline
      & &   & 0.876 & 3K & 8.41s & 0.812\\
     \cline{4-7}
       150 & 20 & $<$1s & & 5K & 10.21s  &0.821\\
     \cline{4-7}
      &  &   & 0.876& 8K & 16.39s & 0.824\\
     \cline{4-7}
        &  &  & 0.876& 10K & 20.16s &0.830\\
     \cline{1-7}
     210 & 20 & $<$2s & 0.865& 3K & 9.49s &0.832\\
     \cline{2-7}
        & 30 & $<$2s &0.82 & 3K & 9.62s &0.790\\
     \hline
\end{tabular}}
\vspace{-1em}
\end{table}
\vspace{-1em}
\section{Related Work}
\label{sec7}
We summarize a few works related to our proposed issues in time series anomaly explanation.
\newline \indent \textbf{Anomaly detection in temporal data}.
Anomaly detection (see \cite{DBLP:journals/csur/ChandolaBK09} as a survey) is a important step in time series management process \cite{DBLP:journals/tkde/JensenPT17}, which
aims to discover unexpected changes in patterns or data values in time series.
Gupta et al. \cite{DBLP:series/synthesis/2014Gupta} summarizes anomaly detection tasks in kinds of temporal data and provide an overview of detection techniques (\emph{e.g.}, statistical techniques, distance-based approaches, classification-based approaches). Autoregression and window moving-average models (\emph{e.g.,} EWMA, ARIMA \cite{DBLP:books/daglib/0070577}) are widely used in outlier points detections \cite{DBLP:journals/tkde/TakeuchiY06}. On the other hand, anomalous subsequences are more challenged to be detected because abnormal behaviors within subsequences are difficult to be distinguished from normal behaviors \cite{DBLP:conf/kdd/ToledanoCBT17}.
Sequence patterns discovery in time series is continuously studied, \emph{i.e.,} \cite{DBLP:conf/vldb/PapadimitriouSF05,DBLP:conf/kdd/Morchen06}.
\cite{DBLP:journals/ml/RebbapragadaPBA09} studies anomalous time series intervals and abnormal subsequences. Further, {high-dimension feature} in time series is taken into account for effectiveness improvement in anomaly detection methods \cite{DBLP:journals/pr/ErfaniRKL16,DBLP:journals/tsmc/LiuLLZ18}.
 \newline \indent As anomaly explanation problems in temporal data have been brought to attention in both research and applications of IoT. Technological breakthroughs are still in demand in developing effective anomaly explanation approaches.
\newline \indent \textbf{Rule-based temporal data cleaning}. Data cleaning and repairing is of great importance in data preprocessing. With the rise of temporal data mining, effective cleaning on temporal data is gaining attention according to its valuable temporal information. Ihab F. Ilyas and Xu Chu give an overview of the end-to-end data cleaning process including error detection and repair methods in \cite{DBLP:books/acm/IlyasC19}.
Both statistical-based \cite{DBLP:conf/sigmod/YakoutBE13,Zhang2016Sequential} and constraints-based \cite{DBLP:journals/pvldb/GolabKKSS09,DBLP:conf/sigmod/SongZWY15} cleaning are widely applied in temporal date quality improvement.
\cite{DBLP:journals/pvldb/GolabKKSS09} extends the idea of constraints from dependencies defined on relational database (\emph{e.g}., \textsc{fd}, \textsc{cfd} in \cite{DBLP:series/synthesis/2012Fan}), and proposes sequential dependencies (\textsc{sd}) to describe the semantics of temporal data. Accordingly, speed constraints are developed in sequential data and applied to time series cleaning solutions \cite{DBLP:conf/sigmod/SongZWY15,Zhang2016Sequential}. Causality analysis tries to reason about the responsibility of a source in causing
errors result. Systems like Scorpion \cite{DBLP:journals/pvldb/0002M13} and DBRx \cite{DBLP:conf/sigmod/ChalamallaIOP14} have been developed to compute the causality and responsibility of violations. The DBRx makes explanation discovery erroneous tuples with desirable properties, namely coverage, preciseness, and conciseness. As the existing techniques mostly focus on relational data. We move a step further in anomaly explanation study in temporal data in  this paper. Our work can also complement the state-of-art data cleaning techniques.
\section{Conclusion}
\label{sec8}
We formalized the anomaly explanation problem in multivariate temporal data and construct a self-contained 3-step method to solve the problem. We  identified anomalies as the violations of types of constraints, and devised set-cover-based algorithms to reason the anomaly events with the given knowledge set. Further, we proposed knowledge update methods to improve the knowledge quality and in turn add to the effectiveness of our method. Experiments on real IoT data showed that the proposed method computed high-quality explanation solutions of anomalies.

\bibliographystyle{IEEEtran}
\bibliography{demo2}

\begin{thebibliography}{10}
\providecommand{\url}[1]{#1}
\csname url@samestyle\endcsname
\providecommand{\newblock}{\relax}
\providecommand{\bibinfo}[2]{#2}
\providecommand{\BIBentrySTDinterwordspacing}{\spaceskip=0pt\relax}
\providecommand{\BIBentryALTinterwordstretchfactor}{4}
\providecommand{\BIBentryALTinterwordspacing}{\spaceskip=\fontdimen2\font plus
\BIBentryALTinterwordstretchfactor\fontdimen3\font minus
  \fontdimen4\font\relax}
\providecommand{\BIBforeignlanguage}[2]{{%
\expandafter\ifx\csname l@#1\endcsname\relax
\typeout{** WARNING: IEEEtran.bst: No hyphenation pattern has been}%
\typeout{** loaded for the language `#1'. Using the pattern for}%
\typeout{** the default language instead.}%
\else
\language=\csname l@#1\endcsname
\fi
#2}}
\providecommand{\BIBdecl}{\relax}
\BIBdecl

\bibitem{DBLP:conf/kdd/ToledanoCBT17}
M.~Toledano, I.~Cohen, Y.~Ben{-}Simhon, and I.~Tadeski, ``Real-time anomaly
  detection system for time series at scale,'' in \emph{Proceedings of the
  {KDD} Workshop on Anomaly Detection}, 2017, pp. 56--65.

\bibitem{DBLP:conf/kdd/DasuLS14}
T.~Dasu, J.~M. Loh, and D.~Srivastava, ``Empirical glitch explanations,'' in
  \emph{The 20th {ACM} {SIGKDD} International Conference on Knowledge Discovery
  and Data Mining, {KDD} '14, New York, NY, {USA} - August 24 - 27, 2014},
  2014, pp. 572--581.

\bibitem{DBLP:journals/tkde/TakeuchiY06}
J.~Takeuchi and K.~Yamanishi, ``A unifying framework for detecting outliers and
  change points from time series,'' \emph{{IEEE} Trans. Knowl. Data Eng.},
  vol.~18, no.~4, pp. 482--492, 2006.

\bibitem{DBLP:series/synthesis/2014Gupta}
M.~Gupta, J.~Gao, C.~C. Aggarwal, and J.~Han, \emph{Outlier Detection for
  Temporal Data}, ser. Synthesis Lectures on Data Mining and Knowledge
  Discovery.\hskip 1em plus 0.5em minus 0.4em\relax Morgan {\&} Claypool
  Publishers, 2014.

\bibitem{DBLP:journals/access/WangW20}
X.~Wang and C.~Wang, ``Time series data cleaning: {A} survey,'' \emph{{IEEE}
  Access}, vol.~8, pp. 1866--1881, 2020.

\bibitem{DBLP:journals/debu/DasuDS16}
T.~Dasu, R.~Duan, and D.~Srivastava, ``Data quality for temporal streams,''
  \emph{{IEEE} Data Eng. Bull.}, vol.~39, no.~2, pp. 78--92, 2016.

\bibitem{Braatz2001Fault}
Braatz, L.~H. Chiang, and E.~L.~R. D, ``Fault detection and diagnosis in
  industrial systems,'' 2001.

\bibitem{DBLP:journals/sadm/FujimakiNTSY09}
R.~Fujimaki and T.~N. et~al, ``Mining abnormal patterns from heterogeneous
  time-series with irrelevant features for fault event detection,'' \emph{Stat.
  Anal. Data Min.}, vol.~2, no.~1, pp. 1--17, 2009.

\bibitem{DBLP:conf/sigmod/ChalamallaIOP14}
A.~Chalamalla, I.~F. Ilyas, M.~Ouzzani, and P.~Papotti, ``Descriptive and
  prescriptive data cleaning,'' in \emph{International Conference on Management
  of Data, {SIGMOD} 2014, Snowbird, UT, USA, June 22-27, 2014}.\hskip 1em plus
  0.5em minus 0.4em\relax {ACM}, 2014, pp. 445--456.

\bibitem{DBLP:books/acm/IlyasC19}
\BIBentryALTinterwordspacing
I.~F. Ilyas and X.~Chu, \emph{Data Cleaning}.\hskip 1em plus 0.5em minus
  0.4em\relax {ACM}, 2019. [Online]. Available:
  \url{https://doi.org/10.1145/3310205}
\BIBentrySTDinterwordspacing

\bibitem{DBLP:conf/sigmod/SongZWY15}
S.~Song, A.~Zhang, J.~Wang, and P.~S. Yu, ``{SCREEN:} stream data cleaning
  under speed constraints,'' in \emph{Proceedings of the 2015 {ACM} {SIGMOD}
  International Conference on Management of Data, Melbourne, Victoria,
  Australia, May 31 - June 4, 2015}, pp. 827--841.

\bibitem{DBLP:conf/dasfaa/YinYWHL18}
W.~Yin, T.~Yue, H.~Wang, Y.~Huang, and Y.~Li, ``Time series cleaning under
  variance constraints,'' in \emph{Database Systems for Advanced Applications -
  {DASFAA} 2018 International Workshops}, ser. Lecture Notes in Computer
  Science, vol. 10829.\hskip 1em plus 0.5em minus 0.4em\relax Springer, 2018,
  pp. 108--113.

\bibitem{DBLP:books/daglib/p/Karp10}
R.~M. Karp, ``Reducibility among combinatorial problems,'' in \emph{50 Years of
  Integer Programming 1958-2008 - From the Early Years to the
  State-of-the-Art}.\hskip 1em plus 0.5em minus 0.4em\relax Springer, 2010, pp.
  219--241.

\bibitem{1992Enhancing}
J.~E. Beasley and K.~Jornsten, ``Enhancing an algorithm for set covering
  problems,'' \emph{European Journal of Operational Research}, vol.~58, no.~2,
  pp. 293--300, 1992.

\bibitem{DBLP:journals/pvldb/DingWSLLG19}
X.~Ding, H.~Wang, J.~Su, Z.~Li, J.~Li, and H.~Gao, ``Cleanits: {A} data
  cleaning system for industrial time series,'' \emph{{PVLDB}}, vol.~12,
  no.~12, pp. 1786--1789, 2019.

\bibitem{DBLP:journals/pvldb/0018HQ00MFZ0ZKJ20}
C.~Wang, X.~Huang, and J.~Q. et~al, ``Apache iotdb: Time-series database for
  internet of things,'' \emph{Proc. {VLDB} Endow.}, vol.~13, no.~12, pp.
  2901--2904, 2020.

\bibitem{2010Outliers}
R.~S. Tsay, ``Outliers, level shifts, and variance changes in time series,''
  \emph{Journal of Forecasting}, vol.~7, no.~1, pp. 1--20, 1988.

\bibitem{1998Outliers}
R.~S. Tsay, D.~Pena, and A.~E. Pankratz, ``Outliers in multivariate time
  series,'' \emph{DES - Working Papers. Statistics and Econometrics. WS}, 1998.

\bibitem{DBLP:journals/csur/ChandolaBK09}
V.~Chandola, A.~Banerjee, and V.~Kumar, ``Anomaly detection: {A} survey,''
  \emph{{ACM} Comput. Surv.}, vol.~41, no.~3, pp. 15:1--15:58, 2009.

\bibitem{DBLP:journals/tkde/JensenPT17}
S.~K. Jensen, T.~B. Pedersen, and C.~Thomsen, ``Time series management systems:
  {A} survey,'' \emph{{IEEE} Trans. Knowl. Data Eng.}, vol.~29, no.~11, pp.
  2581--2600, 2017.

\bibitem{DBLP:books/daglib/0070577}
W.~W.~S. Wei, \emph{Time series analysis - univariate and multivariate
  methods}.\hskip 1em plus 0.5em minus 0.4em\relax Addison-Wesley, 1989.

\bibitem{DBLP:conf/vldb/PapadimitriouSF05}
S.~Papadimitriou, J.~Sun, and C.~Faloutsos, ``Streaming pattern discovery in
  multiple time-series,'' in \emph{PVLDB}.

\bibitem{DBLP:conf/kdd/Morchen06}
F.~M{\"{o}}rchen, ``Algorithms for time series knowledge mining,'' in
  \emph{Proceedings of the Twelfth {ACM} {SIGKDD} International Conference on
  Knowledge Discovery and Data Mining, Philadelphia, PA, USA, August 20-23,
  2006}, 2006, pp. 668--673.

\bibitem{DBLP:journals/ml/RebbapragadaPBA09}
U.~Rebbapragada, P.~Protopapas, C.~E. Brodley, and C.~R. Alcock, ``Finding
  anomalous periodic time series,'' \emph{Machine Learning}, vol.~74, no.~3,
  pp. 281--313, 2009.

\bibitem{DBLP:journals/pr/ErfaniRKL16}
S.~M. Erfani, S.~Rajasegarar, S.~Karunasekera, and C.~Leckie,
  ``High-dimensional and large-scale anomaly detection using a linear one-class
  {SVM} with deep learning,'' \emph{Pattern Recognition}, vol.~58, pp.
  121--134, 2016.

\bibitem{DBLP:journals/tsmc/LiuLLZ18}
H.~Liu, X.~Li, J.~Li, and S.~Zhang, ``Efficient outlier detection for
  high-dimensional data,'' \emph{{IEEE} Trans. Systems, Man, and Cybernetics:
  Systems}, vol.~48, no.~12, pp. 2451--2461, 2018.

\bibitem{DBLP:conf/sigmod/YakoutBE13}
M.~Yakout, L.~Berti{-}{\'{E}}quille, and A.~K. Elmagarmid, ``Don't be scared:
  use scalable automatic repairing with maximal likelihood and bounded
  changes,'' in \emph{Proceedings of the {ACM} {SIGMOD} International
  Conference on Management of Data, {SIGMOD} 2013}, pp. 553--564.

\bibitem{Zhang2016Sequential}
A.~Zhang, S.~Song, and J.~Wang, ``Sequential data cleaning: {A} statistical
  approach,'' in \emph{Proceedings of the International Conference on
  Management of Data, {SIGMOD} Conference}, 2016, pp. 909--924.

\bibitem{DBLP:journals/pvldb/GolabKKSS09}
L.~Golab, H.~J. Karloff, F.~Korn, A.~Saha, and D.~Srivastava, ``Sequential
  dependencies,'' \emph{{PVLDB}}, vol.~2, no.~1, pp. 574--585, 2009.

\bibitem{DBLP:series/synthesis/2012Fan}
W.~Fan and F.~Geerts, \emph{Foundations of Data Quality Management}, ser.
  Synthesis Lectures on Data Management.\hskip 1em plus 0.5em minus 0.4em\relax
  Morgan {\&} Claypool Publishers, 2012.

\bibitem{DBLP:journals/pvldb/0002M13}
E.~Wu and S.~Madden, ``Scorpion: Explaining away outliers in aggregate
  queries,'' \emph{Proc. {VLDB} Endow.}, vol.~6, no.~8, pp. 553--564, 2013.

\end{thebibliography}

\end{document}